\documentclass[11pt,a4paper,english]{article}
\usepackage{amssymb,amsmath,graphicx,multicol}
\usepackage{marginnote}
\pdfoutput=1
\usepackage{amsmath}   
\usepackage{blkarray}
 
\usepackage{xcolor} 
\usepackage{verbatim}
\usepackage{tikz}
\usepackage{amsmath, amssymb, amsfonts, mathrsfs}
\usepackage{blkarray}
\usepackage[T1]{fontenc}
\usepackage{graphicx}
\usepackage{amssymb}
\usepackage{esint}
\usepackage{a4wide}
\usepackage{psfrag}
\usepackage{babel}
\usepackage{epstopdf}
\usepackage{color}
\usepackage[
colorlinks=true, 
linkcolor=black,  
citecolor=blue,   
urlcolor=teal    
]{hyperref}
\usepackage{braket}
\usepackage{ytableau}
\usepackage{caption}
\usepackage{subcaption}
\usepackage{bm}
\usepackage{array}
\usepackage[
    style=phys,
    eprint = true
]{biblatex}
\DeclareFieldFormat[article]{title}{\mkbibemph{#1}}
\newcommand{\EQ}{\begin{equation}}
\newcommand{\EN}{\end{equation}}
\newcommand{\be}{\begin{equation}}
\newcommand{\ee}{\end{equation}}
\newcommand{\bea}{\begin{eqnarray}}
\newcommand{\eea}{\end{eqnarray}}
\newcommand{\bi}{\begin{itemize}}

\newcommand{\ei}{\end{itemize}} 
\newcommand{\bc}{\begin{column}{0.50\textwidth}}
\newcommand{\ec}{\end{column}}
\newcommand{\bcs}{\begin{columns}} 
\newcommand{\ecs}{\end{columns}}
\makeatletter
\usepackage{enumerate}
\newcommand{\half}{\frac{1}{2}}

\newcommand{\br}[1]{\left( #1 \right)}
\newcommand{\sbr}[1]{\left[ #1 \right]}

\newcommand{\dd}{\mathrm{d}}

\usepackage{etoolbox} 

\newif\ifappendix
\appendixfalse

\pretocmd{\appendix}{\appendixtrue}{}{}

\makeatletter
\renewcommand{\@seccntformat}[1]{%
  \ifappendix
    Appendix \csname the#1\endcsname:\;
  \else
    \csname the#1\endcsname\quad
  \fi}
\makeatother
\newcommand{\AScomment}[1]{}

\begin{document}

\vspace*{-1.5cm}
\begin{center}
\LARGE
{\bf
Spectral Decimation of Quantum\\
Many-Body Hamiltonians }

\end{center}
\vspace{0.5cm}
\begin{center}
{\large 
Feng He,
Arthur Hutsalyuk,
Giuseppe Mussardo
and 
Andrea Stampiggi
\footnote{astampig@sissa.it, corresponding author}
\vspace{0.9cm}}

{\sl International School for Advanced Studies (SISSA),\\
Via Bonomea 265, 34136 Trieste, Italy\\[2mm]

INFN Sezione di Trieste, via Valerio 2, 34127 Trieste, Italy\\[2mm]
}

\end{center}

\vspace{0.5cm}
\begin{center}
{\bf Abstract}\\[5mm]
\end{center}

We develop a systematic theory of spectral decimation for quantum many-body Hamiltonians and show that it provides a quantitative probe of emergent symmetries in statistically mixed spectra. Building on an analytical description of statistical mixtures, we derive an explicit expression for the size of a characteristic symmetry sector (CSS), defined as the largest subsequence of levels exhibiting non-Poissonian correlations. The CSS dimension is shown to be the size-biased average of the underlying symmetry sectors, establishing a direct link between spectral statistics and Hilbert-space structure. We apply this framework to two paradigmatic settings: Hilbert-space fragmentation and disorder-induced many-body localization (MBL). In fragmented systems, the CSS reproduces the mixture prediction and isolates correlated subsectors even when the full spectrum appears nearly Poissonian. In the disordered Heisenberg chain, spectral decimation reveals the gradual emergence of integrability through a shrinking CSS, whose statistics exhibit signatures consistent with local integrals of motion. We introduce a characteristic symmetry entropy (CSE) as a finite-size scaling observable and extract, within accessible system sizes, the crossover exponents. Our results establish spectral decimation as a controlled, unbiased and computationally inexpensive diagnostic of hidden structure in many-body spectra, capable of distinguishing between chaotic dynamics, statistical mixtures, and emergent integrability.

\newpage

\section{Introduction}

One of the central limitations in the study of quantum many-body systems is the exponential growth of the Hilbert space with system size. In the absence of exact analytical techniques -- such as the Jordan–Wigner transformation or the Bethe Ansatz -- numerical methods become indispensable. Over the years, a wide array of numerical approaches has been developed to provide non-perturbative insights, including density-matrix renormalization group algorithms \cite{Density-1,Density-2,Renorm-1,Renorm-2,schollwock2011density}, tensor-network techniques \cite{Tensor-network,ran2020tensor,orus2019tensor,silvi2019tensor}, machine-learning-based methods \cite{neural-I,neural-II,neural1,neural2,neural3,neural4,neural5,medvidovic2024neural,du2025artificial} -- see also \cite{Scardicchio_NN} for the first-ever application of neural networks to quantum disordered systems, -- and more recently spectral decimation algorithms \cite{statisticalsignatures}. 

Among the basic features of a quantum many-body Hamiltonian is its energy spectrum. Yet, despite its apparent simplicity, the spectrum contains a remarkably rich amount of information, and overall, it determines the signatures between integrable, non-integrable and non-interacting systems. Even purely numerical spectra -- devoid of analytic input --can reveal essential structural properties of the underlying Hamiltonian. As early as the 1950s, Wigner, Dyson, and Porter \cite{Porter,Porter2,DysonI,DysonII}, and later Berry and Tabor \cite{Berry} introduced the modern viewpoint that the statistics of gaps, i.e. the differences between consecutive energy levels, provide a powerful diagnostic of quantum dynamics. Indeed, through gap statistics, one can probe and quantify the statistical signatures of integrability, chaos, and intermediate regimes. Nowadays, it is well understood that generic, non-integrable quantum systems exhibit Wigner–Dyson level statistics, whereas integrable models, endowed with a nontrivial extensive set of commuting conserved operators, feature independent energy levels -- in other words, Poisson spacings. It is worth recalling that, in the modern theory of out-of-equilibrium systems, the asymptotic regime of non-integrable dynamics can be addressed through the Eigenstate Thermalization Hypothesis (ETH) \cite{Deutsch1,Deutsch2,Srednicki,Rigol1,Rigol2,Moudgalya_2021}, whereas the asymptotic stationary states of integrable models are captured by the framework of the Generalized Gibbs Ensemble (GGE) (see \cite{CalabreseEsslerMussardo2016} and references therein). 

Spectral analysis has recently become a key diagnostic tool in the study of phenomena related to the emergence of symmetry. Such interesting phenomenology can stem from constraints on the Hilbert-space or from the disorder preventing thermalization. The first case is commonly known as Hilbert-space Fragmentation (HSF) \cite{HSFPollman}, which has been studied extensively as a form of ETH or weak-ETH violation. Conventional literature often distinguishes between weak and strong HSF. Overall, fragmentation is diagnosed as the emergence of many disconnected emergent symmetry sectors, ranging from one-dimensional states to exponentially large sub-blocks. The phenomenology of models featuring HSF is vast, ranging from dipole-conserving systems \cite{Khemani_dipole,HSFPollman, Moudgalya_2021, PhysRevX.9.021003,lydzba2024local,classen2025universal} to more exotic pair-flip Hamiltonians \cite{HilbertSpaceFragmentationMotrunich,caha2018pairflipmodel,hart2024exact,stahl2024topologically,Pollmann3}, and we defer the reader to \cite{Moudgalya_2022} for a detailed review on this topic. 

As for the second case, the presence or absence of thermalization in certain disordered quantum many-body systems is a topic of wide research interest. Several groups have focused on the study of many-body localization from different perspectives, for example, by studying the thermalization of observables, e.g., quantum quench dynamics in the many-body localized (MBL) phase \cite{SerbynEtAlQuench2014, Lukin2019_Science}, entanglement growth and scaling properties of eigenstates \cite{AbaninEtAlRMP2019, SerbynEtAlPRL2013,PhysRevLett.124.243601,DumitrescuVasseurPotter,Pietracaprina_2017,BardarsonPollmannMoore,Lukin2019_Science}, renormalization-group approaches to the MBL transition \cite{MorningstarHusePRB2019, DumitrescuEtAl2018, bracci2025}, and spectral statistics and related diagnostics of ergodic–localized crossovers \cite{LuitzLaflorencieAletPRB2015,OganesyanHuse2007}. For a more extensive list of references and contributions to this active field, we refer the reader to broader reviews on many-body localization and thermalization \cite{NandkishoreHuseARCM2015, AletLaflorencieReview2018, SierantEtAlReview2024}.

When considering spectral analysis, it is, however, crucial to recognize that serious obstructions may arise from the presence of emergent ``hidden'' symmetries or, more generally, nontrivial constraints on the Hilbert space. Such constraints, as well as ordinary symmetries, split the Hilbert space into multiple dynamical and disconnected sectors. The global spectrum, being the juxtaposition of the individual sectors, is that of a ``statistical mixture''. Even if each individual sector features level repulsion, i.e. an analogous distribution to the Wigner-Dyson (WD) one appearing in the study of the Gaussian orthogonal ensemble (GOE) in random matrix theory \cite{Mehta1}, the gap probability of the mixture is nonvanishing for arbitrarily small gaps. Consequently, a naive inspection of gap statistics may falsely indicate integrability or suggest the emergence of a non-ergodic phase.

This obstacle was already recognized in Porter’s early work \cite{Porter}, where higher-order gap statistics were introduced to expose deviations from true Poissonian behaviour. Recently, new approaches have been developed to address hidden symmetries and to extract reliable information from many-body spectra --  see \cite{Vernier,statisticalsignatures} and references therein. Fundamentally, the central question can be phrased as follows: given a spectrum whose level gaps appear Poissonian, can one determine whether energy levels are independent, or whether they arise from the superposition of many independent spectral blocks in a statistical mixture?

In \cite{statisticalsignatures}, a new idea based on exact numerical sampling techniques, precursors of Monte-Carlo sampling, was introduced. It was shown that it is possible to discern, from a top-down approach, if a many-body spectrum pertains to a mixture or to a set of independent energy levels -- thereby signaling strong evidence of integrability. Through the application of consecutive removals of Poisson gaps, the spectrum can be decimated to a smaller set of values, whose gap statistics may be GOE-like, as in the case of mixtures, or Poissonian. 

In this paper, we provide a more systematic theory of the spectral decimation of \cite{statisticalsignatures}, and perform an in-depth analysis of statistical mixtures, with applications to constrained and disordered quantum many-body Hamiltonians. 
Starting from the known analytical result for statistical mixtures \cite{Porter2, BerryRobnik, Mehta1}, we derive an approximation for the gap statistics to quantify the fraction of uncorrelated Poisson spacings and the set of correlated ones. We call the latter the ``characteristic symmetry sector'' (CSS), and derive a formula for its size: it is the size-biased average of the individual component dimensions, $d_{\text{out}} = \sum_i d_i^2/d$. Each sub-block contributes to the CSS proportionally to its density of states $d_i/d$, reflecting the probability that two randomly selected levels in the mixture belong to the same block. In this way, spectral decimation dynamically isolates the statistically dominant symmetry sector even when the full spectrum appears Poissonian.

We also present several concrete and illustrative applications of this method; in particular, our focus is on two phenomena:
\begin{enumerate}
    \item 
{\bf Hilbert-Space Fragmentation (HSF)} \cite{HSFPollman,Pollmann1,Pollmann2,Moudgalya_2022}.
HSF occurs when the Hilbert space dynamically decomposes into disjoint sectors due to the presence of constraints, understood as an emergent commutant algebra \cite{HilbertSpaceFragmentationMotrunich}. Importantly, HSF provides an extreme stress-test for gap diagnostics, as it can involve exponentially many sectors, difficult to deduce from a naive study of the Hamiltonian since it generates a large Poisson contribution to the mixture. HSF thus offers a natural arena in which spectral decimation can distinguish genuinely non-integrable dynamics from a deceptively integrable-looking spectrum.

\item {\bf Many-Body Localization (MBL)} \cite{PhysRevLett.95.206603,BaskoAltshuler,BaskoAleinerAltshuler2006,PhysRevB.76.052203,OganesyanHuse2007,huse}.
MBL is a phenomenon in which strongly disordered interacting systems undergo a transition to a phase\footnote{Whether this is a true phase or merely a finite-size MBL regime, with ETH restored in the thermodynamic limit \cite{Vidmar1,PhysRevB.103.024203,PhysRevB.102.064207,LuitzLaflorencieAletPRB2015,PhysRevLett.118.196801,Panda_2019,PhysRevB.105.174205,PhysRevE.104.054105,PhysRevLett.115.187201}, depends on the specific model. Here, we focus on cases where the existence of a bona fide MBL phase is established.} with emergent integrability diagnosed through the presence of independent energy levels,  typically through the statistics of the $r$-ratio \cite{OganesyanHuse2007}. Statistical decimation provides a more refined diagnostic, imputing the origin of emergent integrability not to genuine independent levels, but to those of a disorder-dependent statistical mixture. This phenomenon is well-captured by the dimension of the CSS and quantified by the ``characteristic symmetry entropy'' (CSE) as a function of both the disorder strength and system size.
\end{enumerate}

The paper is organized as follows. Section~\ref{s_mixtures} defines statistical mixtures, recalls the exact gap statistics and derives the small-gap approximation, from which the dimension of the CSS is obtained. Section~\ref{s_decimation} provides a high-level description of the spectral decimation algorithm. Section~\ref{s_applications} covers the application of the spectral decimation to HSF and MBL systems. Our conclusions are gathered in Section~\ref{s_conclusions}.

This paper presents several appendices aimed at complementing the results of the main text. Appendix~\ref{a_unfolding} discusses the unfolding procedure utilized in this work, which explicitly preserves the positivity of the energy density. Appendix~\ref{a_details_decimation} discusses the technical implementation of the spectral decimation, and Appendix~\ref{a_numerical_checks} contains the statistical analysis of the decimation applied to statistical mixtures of random matrices.

\section{Statistical Mixtures}\label{s_mixtures}

The simplest model of a statistical mixture is that of a Hamiltonian system, where each energy level belongs to one of $N$ disjoint ``sectors''. In principle, these sectors may be generated by a symmetry with $N$ irreducible representations on the physical Hilbert space $\mathcal{H}$ or by some constraints. Each sector $\mathcal{H}_i$ is of dimension $d_i$, and there are as many energy levels $E_{j}^{(i)}$, distributed according to the density $\rho_i(E)$, which can be found, for instance, through a best-fit approximation, as outlined in Appendix~\ref{a_unfolding}. In practice, one needs the cumulative density of energies $c_{i} (E) = \int_{-\infty}^{E} \dd x \; \rho_i(x) $, from which the ``unfolded energies'' are simply $\epsilon_j^{(i)} = c_i (E_j^{(i)})$ and the $j$-th consecutive (unfolded) gap is $s_j^{(i)} = \epsilon_{j+1}^{(i)} - \epsilon_j^{(i)}$.

For each sub-block, we denote the statistics of the gaps as $p_{i}(s)$. To compare with physical spectra, one requires $p_i(s)$ to be normalizable and to have a finite average $\mu_i = \int_0^\infty \dd s \; s p_i(s) = 1/\overline{\rho}_i$, with $\overline{\rho}_i$ being the average density. Customarily, $p_i(s)$ is compared to a universal distribution of mean and average one: to this aim, it is always possible to rescale the $s_i \to s_i' = s_i \overline{\rho}_i$, such that $\mu_i = 1$. 

As it will be clear later, the Poisson component of a mixture can come from consecutive energy levels of two distinct blocks, or if $p_i(s)$ is Poissonian, from consecutive energies of block $i$. We restrict ourselves here to the assumption that Poisson gaps come only from levels of distinct blocks; thus, we assume each $p_i(s)$ to exhibit level repulsion, modeled through a Brody-like small gap behavior: $p_i(s) \sim s^{\nu_i}$ as $s\sim 0$, for some $\nu_i > 0$. 

A statistical mixture of the $N$ spectra is the union of the energy levels of each sub-block: $\{E\}_{(N)} = \bigcup_j \{E\}_{j}$. We are interested in $p_N(s)$, the probability density function (PDF) of the consecutive gaps of the mixture, which in general has a very different statistics compared to the one of the individual components, since energy levels belonging to different spectra may fall arbitrarily close, contributing to a Poisson component in $p_N(s)$. 
Analytical formulas were derived by Porter and Rosenzweig \cite{Porter2}, Berry and Robnik \cite{BerryRobnik} and Mehta \cite{Mehta1}.
While exact, they involve first and second integrals of the $p_i(s)$'s, which may be difficult to interpret and manipulate. We shall derive an approximate formula holding in the region $s \ll 1$, which takes a particularly simple form. Before doing that, let's recall the main results by \cite{Porter2,BerryRobnik, Mehta1}. 

Denote by $d = \sum_i d$ the size of the statistical mixture, and $\rho_i = d_i/d$ the density of levels of type $i$, i.e. the probability that two consecutive energies belong to the same sub-block $i$. Let $S$ be the total gap before unfolding, so that the average gap is $\overline{S} = 1/\rho$, with $\rho = \sum_i \rho_i$. Let 
\begin{equation}
    P_i(S) = \int_S^\infty \dd \sigma \; p_i(\sigma), \quad 
    E_i(S) = \int_S^\infty \dd \sigma P_i(\sigma)
\end{equation}
be respectively the probability of a gap larger than $S$ (coming from two consecutive levels of the $i$-th spectrum) and the probability that any two consecutive levels have a gap larger than $S$. When evaluated in $S=0$, $P_i(0) = 0$ is the overall normalization and $E_i(S) = 1/\rho_i$ is the average spacing. The PDF of the mixture is then given by:
\begin{equation}\label{e_exact_statistical_mixture}
    p_N(S) = \frac{\dd^2}{\dd S^2} E(S), \quad E(S) = \frac{1}{\rho} \prod_{i=1}^N \rho_i E_i(S).
\end{equation}
The unfolded gap is $s = \rho S$, and $p_N(s) = p_N(S)/\rho$.

We seek an approximate form when $s \ll 1$ and the constituents follow a Brody-like distribution \cite{Brody}
\begin{equation}
    p_{B}(S, \rho', \nu) = \rho' (1+\nu) a s^{\nu} \exp(-a (S \rho')^{1+\nu})
,
\quad 
a = \left[\Gamma\left(\frac{2+\nu}{1+\nu}\right)\right]^{1+\nu},
\end{equation}
which interpolates between the GOE ($\nu = 1$) and Poisson ($\nu = 0$) gap distributions. The Brody-like distribution is properly normalized and of mean $1/\mu'$. For these distributions, one can compute the integrals explicitly
\begin{equation}
    P_B(S, \rho', \nu) = e^{-(S\rho')^{1+\nu} a}, \quad E_B(S, \rho', \nu) = \frac{\Gamma\left(\frac{1}{1+\nu}, (S\rho')^{1+\nu} a \right)}{\rho \Gamma\left(\frac{1}{1+\nu}\right)},
\end{equation}
where $\Gamma(\alpha, z) = \int_z^{\infty} \dd t\; e^{-t} t^{\alpha -1}$ denotes the incomplete gamma function. The small-$S$ expansions
\begin{equation}
    P_B(S, \rho,\nu) \sim 1 - \left(S \rho \Gamma\left(\frac{2+\nu}{1+\nu}\right)\right)^{1+\nu}, \quad E_B(S,\rho, \nu) \sim 1 - s + \frac{\left(S \rho \right)^{2+\nu}}{2+\nu} \Gamma\left(\frac{2+\nu}{1+\nu}\right)^{1+\nu}
\end{equation}  
can be inserted in Eq.~\eqref{e_exact_statistical_mixture} to obtain the approximation
\begin{equation}\label{e_approximate_statistical_mixture}
    p_N(s) \sim  \sum_{i=1}^N \left(\frac{\rho_i}{\rho}\right)^2 p_i\left(\frac{\rho_i}{\rho}s\right) e^{-s(1-\rho_i/\rho)}+ M e^{-s}, \quad M = 1 - \sum_{i=1}^N \left(\frac{\rho_i}{\rho}\right)^2,
\end{equation}
where here $p_i(s)$ is the standardized distribution with average 1. In the presence of one spectrum, this formula reduces to $p_1(s)$, when $\rho = 1$, and in the limit $N\to\infty$, as $\rho_i \sim 1/N$, only the Poisson component survives, irrespective of the $p_i$'s. Concerning the validity of the approximation, even when $N$ is small, it qualitatively describes the exact solution \cite{Porter2, BerryRobnik, Mehta1} in the interval $s \lesssim 0.5$, see, for example, the plots in Appendix~\ref{a_numerical_checks}.

Generally, this Poisson component of the mixture is a fraction $M$ of it, which also corresponds to the value of the PDF near the origin. What remains, of size $(1-M)$ is the correlated component of the mixture, the ``characteristic symmetry sector'' (CSS). The spectral decimation, applied to a mixture, separates the Poisson component from the CSS. Its output, a spectrum of size $d_{\text{out}}$, will then be distributed according to the correlated component of Eq.~\eqref{e_approximate_statistical_mixture}. The size of the CSS extracted through the spectral decimation is a Gaussian random variable with an expectation value 
\begin{equation}\label{e_expectation_dout}
    \mathbb{E}[d_{\text{out}}] = d(1-M) = \sum_i \frac{d_i^2}{d}.
\end{equation}
This is the central result of this paper: $d_{\text{out}}$ is the size of CSS where all energies are statistically dependent. If all spectra have the same dimensions $d_i = d/N$, then it is easy to see that also $\mathbb{E}[d_{\text{out}}] = d/N$. On the other hand, if the densities are not equal, $d_{\text{out}}$ is dominated by the largest sectors. Finally, if the system has uncorrelated energy levels -- which is the case of strongly-interacting integrable systems and free ones -- $\mathbb{E}[d_{\text{out}}] = 1$, since $d_i = 1$. Information about the variance of $d_{\text{out}}$ can be found in Appendix~\ref{a_details_decimation}.
In other words, in the absence of explicit symmetry generators that allow us to sift the energies of the $N$ blocks, the spectrum of the mixture only retains information about a CSS with a ``size-biased'' dimension, or, in other words, by the frequency that two consecutive energies belong to the same sub-block. The CSS is a natural probe of the two-level spectral correlations. However it also influences dynamical quantities which probe long-range correlations, like the spectral form factor. In Appendix~\ref{a_numerical_checks} this latter dynamical quantity is studied varying the size of the CSS.

\section{Spectral Decimation}\label{s_decimation}

From a statistical perspective, Poissonian spacings signal the absence of level repulsion, either because the system is integrable or because levels from different symmetry sectors cross freely. Spectral decimation exploits this distinction by iteratively removing those spacings that are statistically indistinguishable from a Poisson process. Since level repulsion is preserved under this procedure, correlations originating within the same dynamical sector survive the decimation, while accidental inter-sector crossings are progressively eliminated. Repeating this process isolates a characteristic subset of correlated levels, whose size reflects the dominant symmetry structure of the spectrum.

Thus, the purpose of the algorithm is to extract, via consecutive extractions, the Poisson component of a set of initial $d$ unfolded gaps, obtained from the spectrum of a quantum many-body Hamiltonian -- either the full spectrum or a fraction of it. By construction, it removes the largest fraction of Poisson gaps, leaving a set of $d_{\text{out}}$ remainder gaps. Moreover, it is a self-averaging algorithm: if the input are spectra obtained by $n$ different realizations of the same Hamiltonian (for example, one featuring disorder), it is possible to not only decimate individual samples, but also to collect the gaps together, across all samples. The sample obtained by combining the gaps in this way is called a ``pooled sample''. The $d_{\text{out}}$ obtained by averaging the size of the decimated samples is equivalent to the decimated size of the pooled sample divided by $n$. While the decimation of one sample is more intuitive and sometimes the only option, when possible it is always best to pool many realizations together, as larger samples reduce the statistical fluctuations.

Aside from the initial set of gaps $d$, two other inputs are necessary to the algorithm: the first is an ``extraction fraction'' $f$, i.e. the fraction of Poisson gaps which are removed from the current sample. While $f$ is, in general, arbitrary, too small a fraction can lead to systematic underestimates of $d_{\text{out}}$, while, if too large, it can lead to the exact opposite. Appendix~\ref{a_details_decimation} discusses how to properly choose $f$ to ensure the algorithm stability. The second parameter of the algorithm is the halting size $d_{\text{halt}}$, which is associated to a confidence level $\alpha = 1-d_{\text{halt}}/d$. When the sample has been decimated to a size $d_{\text{halt}}$, at least a fraction $\alpha$ of the initial set of gaps is Poisson distributed. $d_{\text{halt}}$ prevents the algorithm from running indefinitely. When pooling $n$ realizations together, the halting number can be set to $d_{\text{halt}} = n$, implying that if the pooled sample has been decimated to $d_{\text{out}} \leq n$, then the individual samples will have sub-blocks of dimension one, on average.

\begin{figure}
    \centering
    \subfloat[]{\includegraphics[width=0.33\linewidth]{
    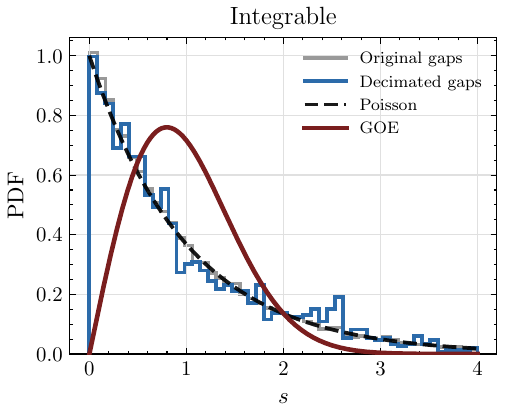
    }}
    \hfill
    \subfloat[]{\includegraphics[width=0.33\linewidth]{
    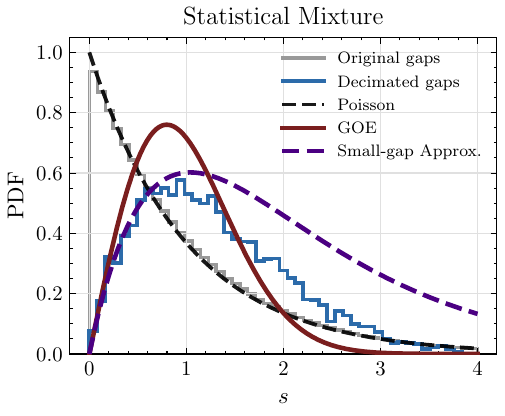
    }}
    \hfill
    \subfloat[]{\includegraphics[width=0.33\linewidth]{
    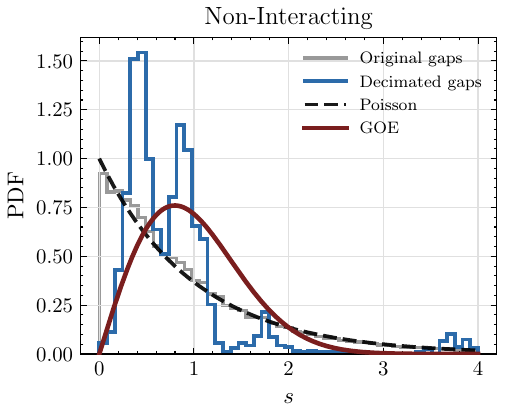
    }}
    \caption{Outcome of the decimation in the case the parent Hamiltonian is a) integrable, b) statistical mixture of chaotic spectra, c) non-interacting. In a) the periodic spin-1/2 XXX model is diagonalized for $L=22$ sites in the $SU(2)$ irreducible representation. The remainder gaps are Poissonian and the decimation succeeds for any $d_{\text{halt}}$. b) The parent Hamiltonian is a mixture of 24 random GOE (Gaussian Orthogonal Ensemble) matrices of same size $d = 10^4$. Here the decimation ends at the size of one GOE block and the gap statistics clearly shows signs of level repulsion. The small-gap expansion found in Section~\ref{s_mixtures} closely follows the decimated gap statistics for $s \lesssim 1$. c) Decimated spectrum of a $L=10$ free fermionic chain with random on-site energies. In particular, they are i.i.d. uniformly in [0,1]. The gap statistics of the decimated gaps presents an excess in the zero-gap probability compared to the pure Poisson case, due to i.i.d. nature of the frequencies.}
    \label{f_statistics_largest_block}
\end{figure}

When the spectrum is that of a mixture, the algorithm ends before meeting the halting threshold: $d_{\text{out}} > d_{\text{halt}}$. In this case, Eq.~\eqref{e_expectation_dout} provides the expectation of $d_{\text{out}}$. On the other hand, if the algorithm ends after meeting the threshold $d_{\text{halt}}$, one concludes that with confidence $\alpha$ the spectrum is composed of uncorrelated energy levels. If the remainder gap statistics is compatible with a Poissonian, this provides a strong hint for the integrability of the quantum many-body system. It may also happen that the remainder gap statistics features distinct peaks: in this case, it is likely that the spectrum is free or weakly-interacting. These outcomes are reported in Fig.~\ref{f_statistics_largest_block}.

Below, give here a short description of the spectral decimation developed in \cite{statisticalsignatures}, referring the reader to Appendix~\ref{a_details_decimation} for implementation details and more details on its statistical properties.

\subsection{Structure of the Algorithm}

Given a sample of size $d$, an iteration $m$ of the decimation algorithm, with extraction fraction $f$ and halting threshold $d_{\text{halt}}$, proceeds as follows:
\begin{enumerate}
    \item the empirical PDF $\pi_j$ of the available ``remainder'' gaps, $\mathcal{R}_m$ of $d_R(m) = (1-f)^m d$ elements, is computed with bin-width 
    \begin{equation}
        \delta(m) = \left(\frac{12}{d_R(m)}\right)^{1/3}.
    \end{equation}
    At the first iteration, $m = 0$, the set of remainder gaps coincides with the original set $\mathcal{R}_0$ of $d$ gaps. The bin-width $\delta$ is chosen according to the Freedman-Diaconis rule for the Poisson PDF \cite{FreedmanDiaconis}, which minimizes the integrated mean squared error between the empirical and true PDF -- see Appendix~\ref{a_FD} for a derivation of the formula.

    \item To the set $\mathcal{R}_m$, one applies rejection-sampling (RS) -- see Appendix~\ref{a_details_decimation} -- with target probability $\tilde{q}(s) = e^{-s}$ and acceptance ratio
    \begin{equation}
        M(m) = \min \left\{ \pi_0 ,  1\right\}.
    \end{equation}
    Here $M(m)$ is the fraction of Poisson gaps, up to order $O(\delta)$, at iteration $m$, estimated from the zero-bin empirical PDF $\pi_0$. One therefore obtains a set $\mathcal{A}_m$ of accepted gaps which follow the Poisson distribution.

    \item A set of $d_E(m) = f d_R(m)$ Poisson gaps is extracted from $\mathcal{A}_m$ uniformly. This yields a subset of $\mathcal{E}_m$ of $d_E$ Poisson gaps at each iteration which is removed from the set of available ones
    \begin{equation}
        \mathcal{R}_{m+1} = \mathcal{R}_m - \mathcal{E}_m.
    \end{equation}

     \item The set $\mathcal{R}_{m+1}$ of $d_R(m+1) = (1-f)^{m+1} d$ remainder gaps can be utilized as the input for the iteration $(m+1)$. 
\end{enumerate}

\paragraph{Halting Conditions}
\begin{enumerate}
    \item [H1.] If after 1., $d_R(m) \leq d_{\text{halt}}$, the algorithm ends and the global spectrum is Poissonian with confidence $\alpha$. 
    \item [H2.] If after 3. and $d_R(m) > d_{\text{halt}}$, the number of accepted gaps, whose expected value is $\mathbb{E}[d_A(m)] = M(m) d_R(m)$, is less than $d_E(m) = f d_R(m)$, the algorithm halts because the requested fraction has not been found. In this case, the algorithm ends and the global spectrum is not Poissonian with confidence $\alpha$.
\end{enumerate}

\paragraph{Generalizations.} The spectral decimation algorithm can be generalized to sample different statistics rather than the one of consecutive gaps. Another choice is the $r$-ratio
\begin{equation}
    r_j = \frac{\min\{E_{j+1} - E_j, E_{j+2} - E_{j+1}\}}{\max\{E_{j+1} - E_j, E_{j+2} - E_{j+1}\}}.
\end{equation}
While the decimation procedure can be applied to different spectral observables (e.g. gaps or $r$-ratios), the statistical mixture approximation leading to Eq.~\eqref{e_expectation_dout} is strictly derived for independent spacings. Since $r$-ratios involve correlated triplets of levels \cite{Vernier}, their mixture structure does not generically obey the same size-biased law. Therefore, the CSS extracted from $r$-ratios needs not coincide quantitatively with the one extracted from gaps, and its derivation is an interesting problem which lies beyond the scope of this paper. We noticed that the CSS size in fragmented Hamiltonians (Section~\ref{s_fragmentation}) and in toy-models of statistical mixtures (Appendix~\ref{a_numerical_checks}) differs from Eq.~\eqref{e_expectation_dout} when computed with the $r$-ratio. On the other hand, for disordered quantum Hamiltonians (Section~\ref{s_MBL}), the CSS size found through the $r$-ratio and gap statistics present the same qualitative behavior, while still featuring quantitative deviations, especially at large disorder couplings. Nevertheless, the $r$-ratio remains a valuable numerical tool as it is inherently independent of the local density of states, bypassing the need for spectral unfolding.

\section{Applications to Quantum Many-Body Hamiltonians}\label{s_applications}

In this Section, we consider realistic quantum many-body Hamiltonians where the Hilbert space presents disconnected sectors, such as in the case of Hilbert space fragmentation, or disorder, which is the case of systems featuring many-body localization. Emergent symmetries, due either to the presence of many Krylov subspaces or local integrals of motion, can cause the global statistics to appear Poissonian. In this section we investigate if the CSS presents the same Poisson statistics, or if it underlies the presence of a statistical mixture, and if its size can inform us about the underlying emergent symmetry.

\subsection{Hilbert-Space Fragmentation}\label{s_fragmentation}

Hilbert-space fragmentation has emerged as a distinct and conceptually rich form of ergodicity breaking in quantum many-body systems, one that is neither rooted in integrability nor driven by quenched disorder \cite{Pollmann1,Pollmann2,Pollmann3,fragment1,fragment2,fragment3,Moudgalya_2022} (see also \cite{HSFPollman,fragment4,PhysRevB.103.L220304,Ising-frag1,Ising-frag2,Q-East,PhysRevB.101.214205,PhysRevLett.124.207602,PhysRevLett.128.196601,PhysRevB.103.214304,PhysRevB.104.155117,
PhysRevResearch.3.033201,10.21468/SciPostPhys.11.4.074,
PhysRevB.103.235133,PhysRevResearch.4.L012003,PhysRevE.104.044106,
10.21468/SciPostPhysCore.4.2.010,
10.21468/SciPostPhys.10.5.099,gp2w-mlkk} for an extensive list of models). In fragmented models, the physically natural (often product-state) basis reveals that the Hamiltonian dynamics does not explore the full Hilbert space but instead decomposes it into an exponential number of dynamically disconnected Krylov subspaces. Unlike conventional symmetry sectors—whose number grows at most polynomially with system size—these Krylov subspaces are generated by microscopic kinetic constraints or conserved dipole-like quantities and can span a broad spectrum of dimensionalities, from finite or logarithmically growing manifolds to subspaces that remain exponentially large. As emphasized in the recent review of fragmentation phenomena \cite{Moudgalya_2022}, this structural decomposition leads to a rich variety of dynamical behaviours: certain subspaces support effectively integrable dynamics with Poissonian spectral statistics, while others exhibit non-integrable, quantum-chaotic behaviour consistent with GOE statistics. Consequently, Hilbert-space fragmentation provides a controlled setting in which integrable and chaotic sectors coexist within a single Hamiltonian, and the global spectral statistics becomes an unreliable indicator of the underlying dynamical properties.

A salient implication of this phenomenon, particularly relevant for our present study, is that conventional diagnostics, such as level-spacing ratios or eigenstate entanglement scaling, can be profoundly obscured by the superposition of spectra originating from multiple dynamically isolated sectors. In weakly fragmented systems, only a measure-zero subset of eigenstates violates ETH, and the majority of sectors behave thermally, often yielding GOE statistics once symmetries are resolved. In strongly fragmented systems, by contrast, the proliferation of small-dimensional Krylov subspaces produces extensive degeneracies and unconventional spectral statistics, typically mimicking Poisson behavior even when the dynamics within individual large sectors remain non-integrable. This mixing problem underscores the need for refined spectroscopic tools capable of disentangling the true dynamical character of each sector. The spectral decimation protocol is precisely designed to meet this challenge: by progressively thinning the spectrum and isolating statistically consistent subsequences, it enables us to diagnose hidden non-integrable sectors in weakly fragmented models and to characterize the emergent integrability structure in strongly fragmented ones. In this sense, Hilbert-space fragmentation provides an ideal testbed for demonstrating the power of spectral decimation in resolving complex ergodicity-breaking mechanisms.

As an example of HSF, we consider the spin-$1/2$ pair-flip (PF) model introduced in \cite{caha2018pairflipmodel,HilbertSpaceFragmentationMotrunich} of the Hamiltonian
\begin{equation}\label{pair_flip}
	H_{\text{PF}} = \sum_{i=1}^{L} \sum_{\alpha, \beta = 1}^2   \left[g_{i,j}^{\alpha,\beta} \ket{\alpha_i\alpha_{i+1}}\bra{\beta_{i}\beta_{i+1}} + h.c\right] + \sum_{i=1}^L \sum_{\alpha = 1}^2 h_i^{\alpha} \ket{\alpha_i}\bra{\alpha_i}\, ,
\end{equation}
where $g_{i,j}^{\alpha,\beta}$ and $h_i^\alpha$ are arbitrary constants.

This model features HSF for any $g_{i,j}^{\alpha,\beta}$ and $h_i^\alpha$ in the computational basis, i.e. the total Hilbert space is decomposed into $(L+1)$ disconnected Krylov spaces, each associated with a $U(1)$ charge. As pointed out in \cite{caha2018pairflipmodel}, each Krylov subspace can be labeled by an ``irreducible string'': for a given state, it is obtained by the deletion of all possible next-nearest-neighbor pairs, going from left to right and repeating if, after a deletion, new pairs arise. For example, if we denote down and up spin states as $1$ and $2$, then the irreducible string for the state $\ket{1221}$ is $\emptyset$, and it is 2 for the state $\ket{22211}$. Each dynamical disconnected Krylov subspace is uniquely characterized by an irreducible string and is spanned by the basis connected  by all allowed pair-flip moves. For example, for the irreducible string $\emptyset$ with $L=4$, one can start from the initial state $\ket{1221} $ and perform all possible pair-flips to generate the basis of the corresponding Krylov subspace, reproducing the action of the Hamiltonian.  One can easily notice that there is no pair flip that allows transitions between states belonging to subspaces with different irreducible strings; therefore, the Hilbert space is fragmented. Alternatively, it is also possible to start from the $2^L$ states of the computational basis, and identify its irreducible strings with a ``pushdown automaton'' \cite{caha2018pairflipmodel}.  

The Hilbert space of this PF model can be expressed as a direct sum of dynamically connected subspaces labeled by an irreducible string $x$, i.e.,
\begin{align}
	\mathcal{H} = \bigoplus \mathcal{H}_x, \quad \mathcal{H}_x= \text{span}\{ \ket{\omega}: \text{irr}(\omega)=x\},
\end{align}
where $\ket{\omega}$ represents the quantum state in Hilbert space, and $x$ denotes the corresponding irreducible string of the state $\ket{\omega}$. In the spin-$1/2$ case, if we denote the total number of 1 and 2 in the irreducible strings as $k = 2M$, it follows that all the possible irreducible strings for a chain with even length $L$ are
\begin{equation}
	\emptyset,\;12,\;21,\;1212,\;2121,\;121212,\;212121,\ldots \;. 
\end{equation}
Each irreducible string $x$ of length $k$ carries a $U(1)$ conserved charge $Q(x)=\sum_{i}^{L-1} (-1)^i (N_i^{(1)}-N_i^{(2)}) = \pm k$, where $N_i^{(1)}$ and $N_i^{(2)}$ count the number of down and up spins in site $i$. If we define an effective magnetization as $M=-Q(x)/2$, then the dimension of the Krylov subspace corresponding to $H_x$ is
\begin{equation}\label{eq_magnetization_sectors}
    D_{L, M} = \binom{L}{\frac{L}{2}- |M|}.
\end{equation}

Therefore, the entire Hilbert space of the PF model is decomposed into $(L+1)$ Krylov subspaces. The largest block is of dimension $D_{L, >}$, which is equal to $D_{L,0}$ for even $L$ and $D_{L, 1/2}$ for odd $L$, respectively. The average block size is $\braket{D_L} = 2^L/(L+1)$.

As the pair-flip model is, to all effects, a statistical mixture, the size of the CSS when all blocks are non-integrable is, according to Eq.~\eqref{e_expectation_dout},
\begin{equation}
    \mathbb{E}[d_{\text{out}}^{\text{PF}}] = 2^{-L} \sum_{M} D_{L,M}^2 = 2^{-L} \binom{2L}{L} \sim \frac{2^L}{\sqrt{\pi L}}.
\end{equation}

As a particular realization of the PF Hamiltonian, we choose the binary random couplings $|g_{i,j}^{\alpha, \beta}| = J(1+\delta_{\alpha,\beta} )/2$ and $h_i^{\alpha}$ to be uniformly i.i.d. in $[-w, w]$ for some disorder strength such that $w/J \ll 1$. With this choice of coefficients, the Krylov subsectors exhibit almost perfect GOE-like level spacing and gap ratio statistics, and therefore serve as an ideal testbed for spectral decimation.

\begin{figure}
    \centering
    \subfloat[]{\includegraphics[width=0.45\linewidth]{
    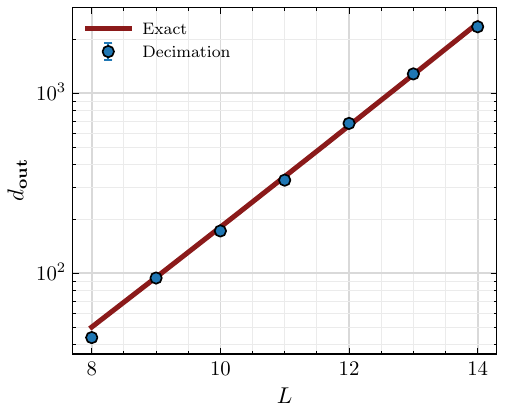
    }}
    \hfill
    \subfloat[]{\includegraphics[width=0.45\linewidth]{
    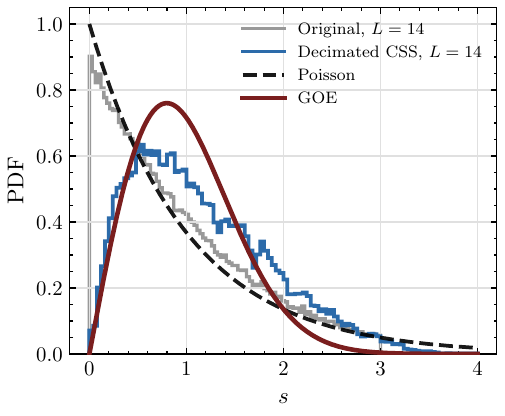
    }}
    
    \caption{Application of spectral decimation to the pair-flip model of Eq.~\eqref{pair_flip}. (a) The decimation output, $d_{\text{out}}$, follows closely the theoretical prediction Eq.~\eqref{e_expectation_dout}. (b) Despite the global distribution of the gaps follows closely the Poisson statistics, the distribution of the remainder gaps, the one surviving the decimation, feature clear sign of level repulsion and therefore the CSS is non-integrable.
    }
    \label{fragmentation}
\end{figure}

For each $L$, we pool several realizations of the PF Hamiltonian together and perform the gap decimation. Fig.~\ref{fragmentation} shows remarkable agreement of the spectral decimation algorithm with Eq.~\eqref{e_expectation_dout}. This pooling approach enlarges linearly the input of the spectral decimation algorithm, and at the same time, it removes the sample-by-sample fluctuations. Therefore, we have demonstrated that the size of the CSS in a generic mixture, Eq.~\eqref{e_expectation_dout}, can be probed through the spectral decimation algorithm, even when HSF occurs.

\subsection{Many-Body Localization}\label{s_MBL}

Many-body localization (MBL) \cite{ PhysRevB.76.052203,BaskoAltshuler,BaskoAleinerAltshuler2006,SierantEtAlReview2024,huse,OganesyanHuse2007,NandkishoreHuseARCM2015, AletLaflorencieReview2018,PhysRevB.90.174202} has emerged as a paradigmatic mechanism of ergodicity breaking in disordered quantum systems, providing a robust alternative to thermalization in highly excited states  \cite{ChoiScience2016,VoskAltmanHusePRL2015,AgarwalAltmanAbanin2017,Agarwal1,randomcouplingXXXscardicchioabanin,PhysRevA.71.012317,PhysRevB.105.224203,PhysRevB.93.060201,Dabholkar:2024jll,Miranda2025LargeDI,PhysRevX.4.011052,Bahri2015,ThieryMuller,PhysRevB.95.155129,SierantDelande,PhysRevB.107.115132,PhysRevB.110.184209,PhysRevB.84.094203,Scardicchio3,10.1098/rsta.2016.0424,DeLuca,PhysRevLett.118.127202,PhysRevResearch.3.L012019,https://doi.org/10.1002/andp.201600360,ImbrieArXiv2016,ImbrieJSP2016}. Unlike conventional isolated many-body systems, whose long-time behavior is governed by the Eigenstate Thermalization Hypothesis (ETH), MBL systems fail to act as their own thermal reservoirs: they retain a persistent memory of their initial conditions and never approach a Gibbs ensemble. This phenomenology
is reflected directly in the structure of the many-body eigenstates: instead of exhibiting volume-law entanglement as predicted by ETH, MBL eigenstates obey area-law scaling even at finite energy density, signaling the emergence of an extensive set of quasi-local integrals of motion (LIOMs or $\ell$-bits) \cite{HuseNandkishoreOganesyanPalPRL2013,ros1,ros2,PhysRevB.91.085425,SerbynPapicAbaninPRL2013,o2016explicit,AbaninEtAlRMP2019, 2408.04338} (see also \cite{PhysRevX.13.011041}). This emergent integrability underpins all aspects of the MBL phase, and it distinguishes MBL sharply from both non-interacting Anderson insulators and integrable models described by Generalised Gibbs Ensembles.

A central challenge in contemporary studies of MBL lies in characterising the transition between thermal and localized phases -- a problem that remains subtle due to finite-size effects, sample-to-sample fluctuations, and the coexistence of thermal and localized regions (avalanches). Conventional diagnostics such as the level-spacing ratio, entanglement entropy, or transport coefficients capture some aspects of this transition, but they are often obscured by Griffiths effects and by the slow approach to the thermodynamic limit. Here, statistical spectroscopy offers a powerful complementary perspective. Because MBL is intrinsically a form of emergent integrability, its onset leaves sharp fingerprints in the spectral statistics: GOE-like distributions characteristic of chaotic dynamics give way to Poisson statistics as LIOMs proliferate. Yet the precise manner in which this crossover occurs—particularly in finite systems where mixed or “rare-region-dominated” spectra are common—requires more refined tools than simple gap-ratio analysis. The spectral-decimation protocol developed in this work is tailored precisely for this purpose: by extracting statistically consistent subsequences of the spectrum, it exposes the latent integrable structure hidden beneath disorder-induced spectral mixing, thereby providing a clean, unbiased probe of the emergent integrability that defines the MBL phase.

As a paradigmatic case of MBL, we analyze the spin-1/2 disordered XXZ model \cite{prosen, huse, alet, CorpsMolinaRelano} of Hamiltonian
\begin{equation}\label{disordered_XXZ}
    H = J\sum_{i=1}^{L} \bm{S}_i \cdot \bm{S}_{i+1} + J(\Delta-1) \sum_{i=1}^{L}S^{Z}_i S^{Z}_{i+1} +  \sum_{i=1}^L h_i S^{Z}_i.
\end{equation}
We set $J = 1$ and $\Delta = 1$. The $h_i$'s are random i.i.d. variables uniformly distributed in $\sbr{-W,W}$, with $W$ the strength of disorder. The model features $U(1)$ symmetry with a generator $\widehat{M} =  \sum_{i} \sigma^Z_i$. Throughout this paper, we focus on the $M=0$ sector for even $L$ and $M=1/2$ for odd $L$. The dimension of the Hilbert space is given by Eq.~\eqref{eq_magnetization_sectors}.

\AScomment{
\footnote{
Using the standard convention for spin-$\half$ operators
\begin{equation*}
    X = 
    \half \sbr{\begin{smallmatrix}
        0 & 1\\
        1 & 0 
    \end{smallmatrix}}, 
    Y = 
    \half \sbr{\begin{smallmatrix}
        0 & -i\\
        i & 0 
    \end{smallmatrix}},
    Z = 
    \half \sbr{\begin{smallmatrix}
        1 & 0\\
        0 & -1 
    \end{smallmatrix}}
\end{equation*}
it is rather easy to show that the permutation operator $P_{i, j}$ is 
\begin{equation*}
    P_{i, j} = 2 \br{ X_i X_j + Y_i Y_j + Z_i Z_j} + \half \id.
\end{equation*}
The Hamiltonian, then reads
\begin{equation*}
    H = \half \sum_{i=1}^{L-1} P_{i, i+1} +  \sum_{i=1}^L h_i Z_i.
\end{equation*}}
}

We now summarize the results of the spectral decimation applied to Eq.~\eqref{disordered_XXZ}, offering both qualitative and quantitative analyses. For each value of $L = \{9, \ldots, 16\}$ we considered at least $N=1000$ samples for each $(W,L)$. Then, for each $(W,L)$ we pool all samples together and perform the decimation on this larger set, resampled $r = 25$ times.

\subsubsection{Qualitative Features of the CSS}

Overall, the CSS remains close to the Hilbert space dimension for $W \lesssim 2$ and then decreases exponentially with an $(W,L)$-dependent slope. Crucially, even at moderate volumes, we observe in the strong disorder region, $W = 10$, that the CSS approaches values on the order of a few percent of $d(L)$, up to $1\%$. In that region, energy levels tend to form small CSS, implying that global spectral correlations are weakening. The operators enabling these emergent symmetries are the LIOMs and the CSS may be interpreted as an estimator of the effective symmetry sectors generated by them. Thus, spectral decimation provides an independent spectral diagnostic consistent with the presence of LIOMs in the disordered XXZ Hamiltonian.

We emphasize that the extracted CSS size is robust under moderate variations of the halting threshold $d_{\text{halt}}$, provided that the extraction fraction $f$ is chosen according to Appendix~\ref{a_details_decimation} so that it does not introduce systematic over- and under-estimates of the decimation output. As explained in Sec.~\ref{s_decimation}, the self-averaging pooling procedure suppresses finite-sample fluctuations, rendering the CSS stable against such variations. We also compared the CSS size obtained using the $r$-ratios instead of the gaps, as an internal test of the consistency of the spectral decimation. Qualitatively, the CSS size presents similar $L$ and $W$ dependence, but their quantitative values differ, as outlined in Section~\ref{s_decimation}.

\begin{figure}
    \centering
    \subfloat[]{\includegraphics[width=0.45\linewidth]{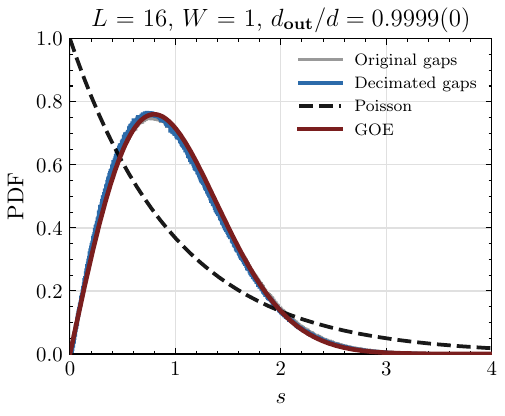}}
    \hfill
    \subfloat[]{\includegraphics[width=0.45\linewidth]{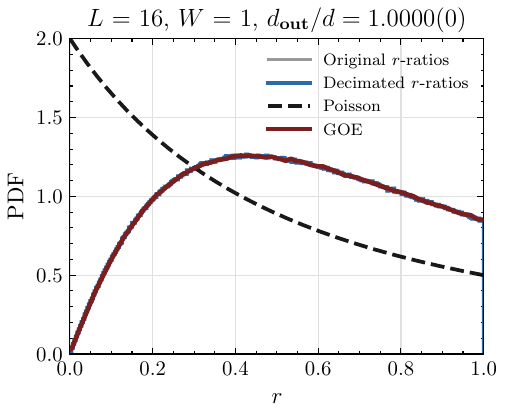}}
    \hfill
    \subfloat[]{\includegraphics[width=0.45\linewidth]{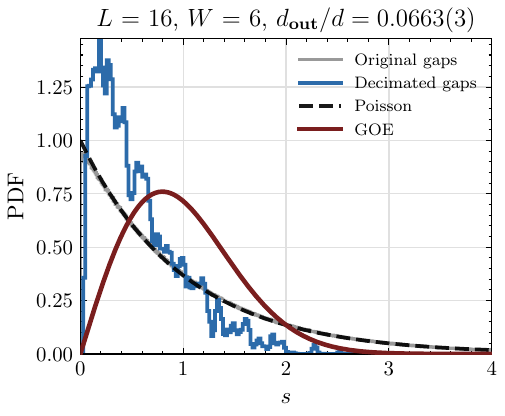}}
    \hfill
    \subfloat[]{\includegraphics[width=0.45\linewidth]{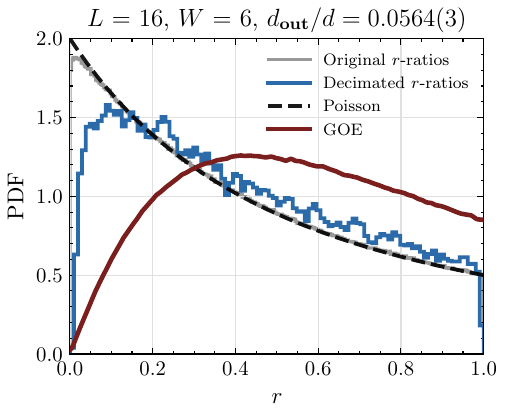}}
    \caption{Histograms of gaps and $r$-ratios of the one-dimensional disordered Heisenberg chain for $L$ = 16 at weak ($W = 0.5$) and strong ($W = 6.5$) disorders, in a energy window of $\Delta E = 90 \%$ of the full spectrum. At weak disorder, one observe excellent agreement of the distribution of gaps (a) and $r$-ratios (b) with the Wigner-Dyson distribution of GOE random matrices. The value of $d_{\text{out}}/d$ is approximately 1 in both cases, the one obtained from the $r$-ratios subject to smaller statistical fluctuations. At strong disorder, on the other hand, while the $r$-ratio statistics (d) seems to be compatible with a semi-Poissonian (which eventually will be restored as $W\to\infty$), the gap statistics (c) clearly deviates from the Poisson one and accumulates towards $s\to0$. Sharp peaks are characteristic of effectively free or weakly interacting spectra, where spectra are regular and energy differences expressible as differences of linear combinations of one-particle energies. This distinction between free and strongly-interacting integrable system statistics is not captured by the $r$-ratio distribution, which is the same. We also observe that the CSS size extracted from r-ratio decimation deviates quantitatively from that of gap decimation, as discussed in Section~\ref{s_decimation}.}
    \label{f_dout_one}
\end{figure}

Fig.~\ref{f_dout_one} shows the underlying distribution of the CSS gaps and $r$-ratios, at weak and strong disorder. We observe that, at weak disorder,  the gap and that of the $r$-ratio statistics are GOE-like. At very large disorder, the leading term of the Hamiltonian is $H_Z = \sum_{i=1}^L h_i S^{Z}_i$, whose spectrum is that of a sum of independent, uniformly i.i.d., $L$ on-site energies. In the limit $L\to \infty$, the statistics of the ``truncated oscillator'' approaches the Poissonian, similarly to the $N\to\infty$ limit of the mixture (see Section~\ref{s_mixtures}). As already pointed out in \cite{Berry}, the level statistics of the truncated oscillator with few frequencies, corresponding in our case to finite $L$, features characteristic peaks due to equidistant energy levels. Moreover, these peaks can not be explained, for example, by a Brody parameter, which only interpolates between GOE and Poisson statistics. The CSS features the same phenomenology at strong disorder. While this is evident from the gap statistics (especially if compared to the free-fermion statistics Fig~\ref{f_statistics_largest_block}(c)), we notice that this feature is absent in the one of the $r$-ratio, which mildly deviates from $p(r) = 2/(1+r)^2$. 

The analysis of the CSS statistics provides new insight into the `emergent integrability' of the one-dimensional disordered XXZ Hamiltonian. It should not be understood as the emergence of generic integrable dynamics, like the one of the same model in absence of disorder, the strongly interacting XXX chain, but of an asymptotically free one, like that of weakly-interacting fermions.

\subsection{Characteristic Symmetry Entropy}

For any $(W,L)$, the CSS size $d_{\text{out}}(W, L)$ obtained by running multiple times (resampling) the gap decimation of the pooled samples varies exponentially with $W$ and $L$. Therefore, the CSS size cannot be used as a sensible observable for a finite-size scaling (FSS) analysis. 

 We start by noticing that, as the asymptotics of the Hilbert space dimension is
\begin{equation}
    \log d(L) \sim (\log 2) L - \frac{1}{2} \log L + \frac{1}{2} \log \frac{2}{\pi},
\end{equation}
one can consider $\frac{1}{L} \log d(L)$ as an intensive quantity, up to $O((\log L)/L)$ corrections. Therefore, we study the ratio of the original size and the one of the CSS dimension. This quantity expresses the effective number of blocks of dimension of the CSS -- assuming them uniformly. The logarithm of this ratio can be rescaled to define the intensive `characteristic symmetry entropy' (CSE):
\begin{equation}
    \Sigma(W,L) = \frac{1}{L \log 2} \left(\log d(L) - \log {d_{\text{out}}(W,L)}\right),
\end{equation}
where $d_{\text{out}}(W,L)$ is the average of the output of the spectral decimation. Vanishing $\Sigma(W,L)$ corresponds to fully chaotic (GOE-like) behavior, while larger values indicate increasing effective fragmentation.

\begin{figure}
    \centering
    \subfloat[]{\includegraphics[width=0.45\linewidth]{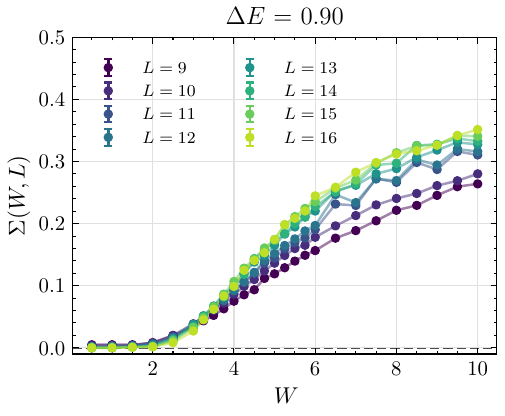}}
    \hfill
    \subfloat[]{\includegraphics[width=0.45\linewidth]{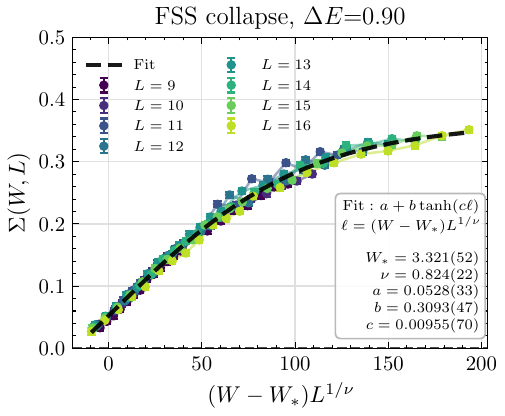}}
    \caption{(a) Plot of the CSE over a large window of the spectrum ($\Delta E = 0.9$), varying the disorder strength $W$, for different volumes $L$. For $W\lesssim 3$, the CSE remains close to 0, indicating that the system is close to GOE, while it increases for larger $W$'s. For the disorder strengths and volumes studied, the CSE does not approach unity, even though the CSS may be as low as $1\%$ of the Hilbert space dimension. As such, the CSE is a sensible probe of the fine spectral structure of disordered quantum Hamiltonians. (b) FSS collapse of the various $\Sigma(W,L)$ curves. Fitted parameters, as well as the fitting form, are reported in the box on the bottom right.}
    \label{f_CSE_FSS}
\end{figure}

We notice that decimations performed with gaps or with $r$-ratios produce qualitatively similar CSE, for the disordered XXZ Hamiltonian. We also observe that it never saturates to unity for the system sizes and disorder parameters considered, reaching a maximum value of approximately $0.4$, see Fig.~\ref{f_CSE_FSS}(a). This  immediately puts into question if the system sizes under consideration represent a legitimate thermodynamic regime, where true MBL scaling dominates. This issue can not be resolved by exact diagonalization alone, and it can be clarified by a renormalization-group study of the CSS, a task which we will not pursue in this current work.

\subsubsection{Finite-Size Scaling across Multiple Energy Windows}

Below, we present a FSS analysis of the CSE. Its purpose is not to draw conclusions on the physics of MBL, rather to show that the CSE is an observable which has FSS collapse. Indeed, the validity of this analysis is based on a phenomenological ansatz and not on a renormalization-group scaling form. Moreover, it is likely that the system sizes utilized are not representative of a true thermodynamic phase, but of a crossover regime. For these reasons, the following analysis will only be indicative rather than conclusive.

While conventional studies of MBL focus on thermal states in a small energy window $\Delta E$, to compensate for the limited system sizes at our disposal, we will also consider wider windows, ranging from the bulk to the full spectrum, where the spectral decimation becomes more statistically accurate. As customary, we introduce a scaling variable 
\begin{equation}
    \ell = L^{1/\nu} (W-W_*),
\end{equation}
where conventionally $W_*$ is the critical disorder parameter and $\nu$ is the dynamical exponent characterizing the divergence of the correlation length: $\xi \sim |W-W_*|^{-\nu}$. It is also worth noting that this ansatz may, in principle, be wrong, as renormalization-group studies \cite{DumitrescuEtAl2018, MorningstarHusePRB2019, PhysRevB.102.125134} argue that the transition may be of the Kosterlitz-Thouless (KT) type, where $\nu = \infty$. Therefore, the scaling parameters should be interpreted as effective crossover quantities rather than true thermodynamical exponents. We tried fitting with both a KT law, $\log \xi \sim (-\sqrt{W-W_*})$, and a power-law one, but found that the KT FSS ansatz does not provide a collapse for the system sizes under study. 

\begin{figure}
    \centering
    \subfloat[]{\includegraphics[width=0.45\linewidth]{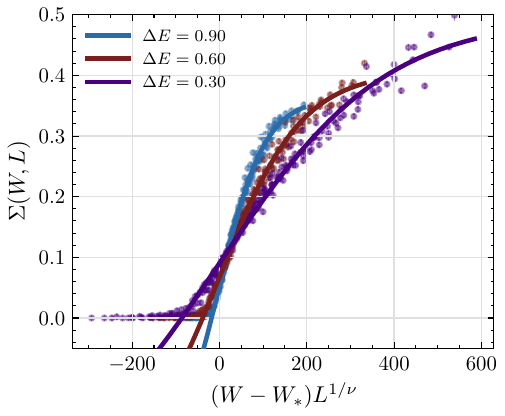}}
    \hfill
    \subfloat[]{\includegraphics[width=0.45\linewidth]{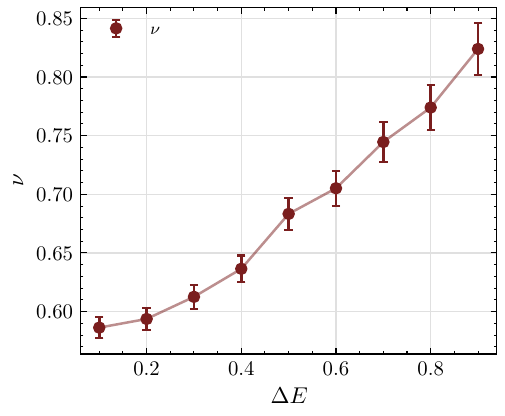}}
    \hfill
    \subfloat[]{\includegraphics[width=0.45\linewidth]{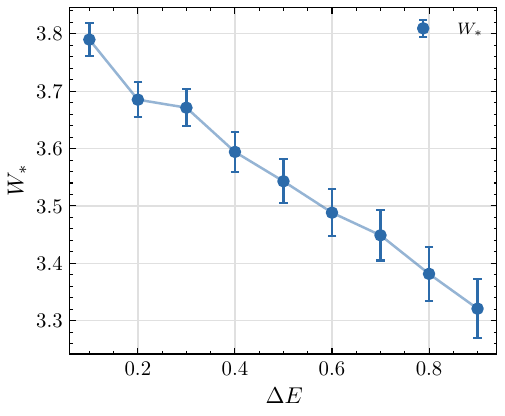}}
    \hfill
    \subfloat[]{\includegraphics[width=0.45\linewidth]{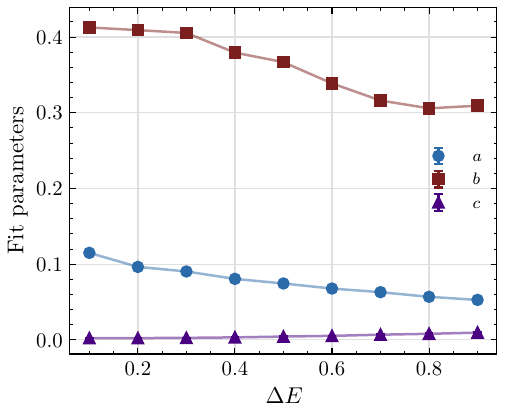}}
    \caption{Results of finite-size scaling (FSS) of the characteristic symmetry entropy (CSE). (a) the FSS of the CSE is reported for few energy windows $\Delta E$, showing that as $\Delta E$ increases, the steepness of the curves, dictated by the critical exponent $\nu$ diminishes. (b) The same critical exponent $\nu$ is plotted against $\Delta E$ for the energy windows studied. (c) Similar plot for $W_*$. Finally, the fit constants are varying monotonically with the energy window (d).}
    \label{f_MBL_plots}
\end{figure}

Overall, the CSE presents three qualitatively distinct regions: for weak disorder, $W\lesssim 3$, it is qualitatively vanishing, then it undergoes a crossover, $3 \lesssim W \lesssim 6$, with an $L$-dependent slope towards a saturation region, $W\gtrsim 10$. This suggests that the FSS form may be a sigmoid or an hyperbolic tangent. Indeed, the last option offers a collapse of the $\Sigma$ versus $W$ at different $L$'s:
\begin{equation}
    \Sigma(W,L) = a + b \tanh(c\ell),
\end{equation}
where $a$, $b$ and $c$ are numerical constants. The FSS ansatz is robust in the disorder window $3\leq W\leq 10$, offering a collapse of the CSE, see Fig.~\ref{f_CSE_FSS}(b). Varying the energy window, we observe that the estimated critical point lies in the range $3 \lesssim W_* \lesssim 4$, and that the critical exponent $\nu < 1$, see Fig.~\ref{f_MBL_plots}.

While the value of $W_*$ for small $\Delta E$ agrees with other estimates from exact diagonalization studies \cite{alet}, the value of $\nu$ violates the Harris bound ($\nu > 2/D$ where $D$ is the spatial dimension). Nevertheless, among the ansatzes tested, the power-law form provides the best collapse of the available data, and we report in Fig.~\ref{f_MBL_plots} its parameters.

Nonetheless, the CSE can be used as an observable easily accessible from numerical data. Moreover, it can be computed either from the gaps, which require unfolding, or the $r$-ratios, which gauge three-level correlations but do not require fitting of the energy density. Thus, it can be considered as a complementary probe of eigenvalue correlations, like the average $r$-ratio. The CSE is, moreover, consistent with a power-law FSS ansatz, which brings it to a volume-independent collapse.

\section{Conclusions}\label{s_conclusions}

In this paper, we applied the spectral decimation procedure \cite{statisticalsignatures} to one-dimensional quantum many-body Hamiltonians featuring nontrivial forms of emergent symmetries. We argued that the underlying framework in which they can be understood is that of statistical mixtures, for which we derived the size of a characteristic symmetry sector (CSS). The CSS is the largest sector with non-Poissonian statistics, characterized either by level repulsion or by isolated peaks, and its size is a size-biased average of the symmetry sectors present in the Hamiltonian.

For systems featuring Hilbert space fragmentation, in particular, the pair-flip model, we showed that the CSS follows the relation predicted by the statistical mixture approximation and demonstrated that, while the overall global statistics may be close to Poissonian, the statistics of the CSS clearly features level repulsion.

For systems undergoing many-body localization, we focused on the one-dimensional disordered Heisenberg chain and showed that the CSS can be used as a probe of the emergent symmetries generated by the local integrals of motion. We clarified the nature of the emergent integrability by looking at the statistics of the CSS at large disorder, which shows clear peaks reminiscent of those of weakly-interacting or almost-free systems. The size of the CSS can be transformed into a novel order parameter, the characteristic symmetry entropy (CSE), which is computationally inexpensive since it is derived from the Hamiltonian spectrum. Alike other observables, such as the $r$-ratio \cite{OganesyanHuse2007,De_2022,SierantZakrzewski2} or the zeroes of the renormalization group $\beta$-function  \cite{GoremykinaVasseurSerbyn,MorningstarHusePRB2019,ZhangZhaoDevakulHuse,ModakNag,VoskAltman2,PotterVasseurParameswaran,DumitrescuEtAl2018,MBL-ETH}, it features universal finite-size scaling collapse. An interesting future direction is to derive the CSE finite-size scaling form from an explicit renormalization-group description of LIOM formation, and clarify the nature of the transition, if of the Kosterlitz–Thouless type or governed by a different infinite-randomness scenario.

In conclusion, we demonstrated that the CSS and the CSE can be used both as qualitative and quantitative probes of the statistical properties of many-body spectra, able to offer guidance in discriminating between strongly-interacting integrable systems, free ones and statistical mixtures of non-integrable components. Moreover, as spectral decimation operates directly on spectral statistics without relying on entanglement or dynamical probes, it provides a computationally inexpensive and broadly applicable diagnostic for emergent integrability.

\section*{Code Availability}

The decimation algorithm is public on \verb|https://github.com/astampig/spectral_decimation| \cite{stampiggi_spectral_decimation_2026}.

\section*{Acknowledgments}

The authors are thankful to Antonello Scardicchio for insightful discussions. AH is grateful to Lenard Zadnik for valuable explanations. We acknowledge financial support from PNRR M4C2I1.3 PE\_00000023\_NQSTI Grant, Spoke A2, funded by NextGenerationEU. AS's research was supported in part by grant NSF PHY-2309135 to the Kavli Institute for Theoretical Physics (KITP).
\appendix

\section{Unfolding Procedure with Bernstein Polynomials}\label{a_unfolding}

In this Appendix, we describe the unfolding procedure used to obtain consecutive spacings from the energy spectrum, defined as
\begin{equation}
    s_i = e_i - e_{i-1}, \quad e_i = \int_{-\infty}^{E_i} \dd E \; \rho(E),
\end{equation}
where $\rho$ is the density of energy. Fitting directly $\rho$ is impractical here, first because it is subject to fluctuations, especially for very small system sizes. Second, we recognize that $e_i$ is the cumulative energy density (CDE), which is a much better quantity to fit, as it is monotonic.

Thus, we seek to fit the CDE, whose empirical values are
\begin{equation}
    \tilde{c}_i = \sum_{j = 1}^{d} \theta(E_i - E_j),
\end{equation}
i.e. it counts the number of energies less than or equal to $E_i$. We notice that, to obtain $s_i$, we can rescale the energies in the region $[0,1]$:
\begin{equation}
    \tilde{E}_i = \frac{E_i - E_1}{E_d - E_1}.
\end{equation}
Then, the problem is reduced to fit the data $(i/d, \tilde{c}_i)$ in the interval [0,1].

The most important property that the fit must preserve is the monotonicity of the $\tilde{c}$'s, which guaranties that $\rho$ will always be positive. A linear polynomial fit with basis functions $x^n$ does not provide generic monotonic $\tilde{c}$'s. In [0,1] there exists a basis of polynomials, called Bernstein polynomials $b_{\nu,n}(x)$:
\begin{equation}
    b_{\nu,n}(x) = \binom{n}{\nu} x^\nu (1-x)^{n-\nu},\; \nu \in \{0, 1, \ldots, n\},
\end{equation}
which is of order $n$.

Bernstein polynomials satisfy:
\begin{equation}
    \frac{\dd}{\dd x}b_{\nu,n}(x) = n \left[b_{\nu-1, n-1}(x) - b_{\nu,n-1}\right].
\end{equation}
If then one considers a generic polynomial approximation of degree $n$ of a monotonic function $f(x)$, i.e.
\begin{equation}
    f_{n}(x) = \sum_{\nu = 0}^n \gamma_{\nu} b_{\nu,n}(x),
\end{equation}
it is easy to show that 
\begin{equation}
    \gamma_0 \leq \gamma_1 \leq \ldots \leq \gamma_n \to \frac{\dd}{\dd x} f_{n}(x) \geq 0.
\end{equation}
Introducing auxiliary constants $\beta_i \geq 0$ such that $\gamma_k = \sum_{i = 0}^k \beta_i$, one can easily perform a constrained linear least-squares fit to obtain the $\beta_i$'s and therefore obtain the best monotonic approximation of a monotonic $f(x)$ through a polynomial of degree $n$. This is exactly the strategy we use to obtain the CDE and, as a byproduct, consecutive gaps ensuring the fitted density is positive.

\section{Details of the Decimation Algorithm}\label{a_details_decimation}
This appendix details certain technical aspects of spectral decimation, which complement the discussion in the main text. Here, we discuss in detail the rejection sampling (RS) algorithm, the optimal bin-width, the choice of the extraction fraction $f$ and sources of systematic errors. Finally, we offer a theoretical prediction of the variance of the decimation algorithm, as it is repeated on the same spectrum many times (resample).

\subsection{Rejection Sampling}

The RS algorithm is an exact simulation technique to  construct a prescribed target distribution $\tilde q(s)$
from samples 
$\{s\}$  drawn according to an original distribution 
$q(s)$. The construction proceeds by generating an empirical distribution $\pi(s)$, which converges to $\tilde q(s)$
in the limit of infinitely many samples. Such a reconstruction is feasible only when the target distribution is fully enveloped within the original one. In particular, if $\tilde q(s)$ is defined over $\tilde d$ 
 elements and $q(s)$ over $d > \tilde d$, then $\tilde q(s)$ 
can be reproduced exactly, without any distortion, provided that
$\tilde q(s) \tilde d \leq q(s) d$, 
for all $s\geq 0$.

From the envelope principle, it follows that only a ratio 
\begin{equation}
    M = \min_s \frac{\tilde{d} \tilde{q}(s)}{d q(s)}
\end{equation}
of the original $d$ gaps can reproduce exactly the distribution $\tilde{q}(s)$.

The set of accepted values $\mathcal{A}$ is that for which 
\begin{equation}
    \mathcal{A} = \left\{ s_j \in \{s_1,\ldots, s_d\}, \;\text{such that}\; u_j q(s_j) \leq M \tilde{q}(s_j), \; u_j \sim \operatorname{Unif}(0,1) \right\}.
\end{equation}
In other words, for each $s_j$, a uniform i.i.d. number in $[0,1]$, $u_j$, is extracted and the gap $s_j$ is conditionally accepted according to the inequality $u_j \leq \alpha(s_j)$, where $\alpha(s) =  \min \{M \tilde{q}(s)/q(s), 1\}$ is the acceptance probability.

It follows that $M$ becomes the ``acceptance rate'' of the RS. Indeed, for each $s_j$, the probability of acceptance is $\alpha(s_j)$ and therefore the total number of accepted values is
\begin{equation}
    \mathbb{E}[d_A] = d \int_0^\infty \dd s\; q(s)\alpha(s) = M d.
\end{equation}
Thus, in one realization of the RS, the number of accepted values $d_A$ is the sum of $d$ Bernoulli variables, accepted with probability $M$. The number of accepted counts, $d_A$, has therefore mean $\mathbb{E}[d_A]$ and variance $\sigma_A^2 = d M (1-M)$. Compactly, we write: $d_A \sim \operatorname{Bern}(d, M)$.

\subsection{Optimal Bin-Width}\label{a_FD}

A central problem in numerical estimation of continuous PDFs is the estimation of the proper histogram bin-width. If the bin-width is too large, the histogram fails to capture the intrinsic variations of the PDF, while if too small, bin counts fluctuations fail to reproduce a smooth curve. For these reasons, choosing an appropriate bin-width is not only of qualitative, but also of quantitative importance.

Let us consider a sample of $d$ i.i.d. elements drawn from the PDF $p(s)$. For its discrete representation, we homogeneously partition the interval $s\in[0, \infty)$ in bins of with $\delta$, such that bin $j$ comprises values $s \in [\delta j, \delta(j + 1)]$. Let $d_j$ be the number of samples falling in bin $j$. The empirical histogram PDF is 
\begin{equation}
    \tilde{\pi}_j = \frac{d_j}{\delta d},
\end{equation}
where $\tilde{\pi}_j$ is in principle a random variable, because $d_j$ is the sum of Bernoulli counts. Indeed, the expectation of $\tilde{\pi}_j$ is the mean of the continuous PDF over the bin:
\begin{equation}
    \mathbb{E}[\tilde{\pi}_j] = \pi_j =\frac{1}{\delta}\int_{j\delta}^{(j+1)\delta} \dd s \; p(s).
\end{equation}
Its variance, on the other hand, is, for $d_j \gg 1$,
\begin{equation}
    \operatorname{var}[\tilde{\pi}_j] = \frac{\operatorname{var}[d_j]}{(\delta d)^2} = \frac{d_j}{(\delta d)^2}.
\end{equation}

The optimal bin-width is that which minimizes the fluctuations between $\tilde{\pi}_j$, $\pi_j$ and $p(s)$ in each bin, over the whole histogram. This quantity is called the integrated mean squared error (IMSE), defined as
\begin{equation}
    \operatorname{IMSE} = \sum_{j} \delta \left(\tilde{\pi}_j - \pi_j \right)^2+\sum_{j} \int_{j\delta}^{(j+1)\delta} \dd s \; \left(p(s) - \pi_j \right)^2.
\end{equation}
The reason for the second term as follows. Denoting as $s_j = \delta(j + 1/2)$ the midpoint value, $\pi_j$ approximates $p(s_j)$ to an order $\delta^2$, which is subleading in the IMSE.

We evaluate the two terms separately. The first is quantified by the variance of $\tilde{\pi}_j$, which comes from the Bernoulli distribution of $d_j$. When summed over bins, it gives:
\begin{equation}
    \sigma^2_\text{hist} = \sum_j \delta \frac{d_j}{(\delta d)^2} = \frac{1}{\delta d}.
\end{equation}
On the other hand, the variance of the integral mean is computed by expanding $q(s)$ around the midpoint $\delta_j$:
\begin{equation}
    \sigma^2_\text{mean}=\frac{\mathcal{I}_{1,2}[p]}{12}
        \delta^2, \quad \mathcal{I}_{n,m}[p] = \int_0^\infty \dd \sigma\; \left(\frac{\dd^n p(\sigma)}{\dd \sigma^n }\right)^m.
\end{equation}
The minimum of their sum yields the Freedman-Diaconis rule \cite{FreedmanDiaconis}:
\begin{equation}
    \delta = \left(\frac{6}{\mathcal{I}_{1,2}[p] d}\right)^{1/3}.
\end{equation}
For the Poisson distribution, $\mathcal{I}_{1,2} = 1/2$.

\subsection{Choice of the Extraction fraction} 

The extraction fraction $f$ is a free parameter of the spectral decimation algorithm. However, if $f$ is chosen too small, gaps can be deemed Poissonian even if the underlying spectrum is that of a random matrix, like a GOE. Therefore, for a spectrum of $d$ elements, one fixes $f$ by looking at the first-bin value of a GOE.
    
For a GOE matrix, the bulk gap PDF is well-approximated by the Wigner-Dyson surmise $p_{\text{GOE}}(s) = \frac{\pi}{2} s e^{-s^2 \pi/4}$. However, it is exact only for $2\times2$ GOE matrices, but not for arbitrary size, see \cite{Mehta1} for a derivation of lower and upper bounds on the GOE CDF. This discrepancy is noticeable at small $s$, where the true GOE PDF behaves as $\pi s/6$; hence, the value of the zero-bin PDF is 
\begin{equation}
    \pi_{0, \text{GOE}}(\delta) = \frac{\pi^2}{12}\delta,
\end{equation}
and its variance is 
\begin{equation}
    \sigma_{\text{GOE}, 0}^2 = \frac{\pi_{0, \text{GOE}} ( 1- \delta \pi_{0, \text{GOE}})}{d\delta}.
\end{equation}

In practice, one chooses $f = \pi_{0, \text{GOE}}(\delta) + z_q \sigma_{\text{GOE}, 0}$, such that if, at a given iteration, a small-gap probability is compatible with a GOE, then the algorithm halts. Assuming $d$ is sufficiently large, $z_q$ is the probit of a normal distribution $\mathcal{N}(0,1)$ of a certain quantile $q$. If at a certain iteration the algorithm ends, it means that the underlying remainder distribution is GOE-like with a zero-bin value compatible with $\pi_{0, \text{GOE}}$ within confidence $q$. For example, when $q = 0.95$, $z_q = 1.96$. This choice of $f$ guaranties that any GOE gap distribution will result in the decimation algorithm halting, thereby returning its actual size.

In the case of systems with randomness, we utilized $f = \pi_{0, \text{GOE}}(\delta) + 1.96 \sigma_{\text{GOE}, 0}$ for sample-by-sample decimations and $f = \pi_{0, \text{GOE}}(\delta)$ when samples are pooled together, as fluctuations are averaged in each bin.

\subsection{Sources of systematic error} 

The choices of a finite sample $d$ and resolution $\delta$ introduce numerical artifacts, independent of $f$. This is especially relevant when the initial gaps are i.i.d. Poisson ones, as this case is a benchmark of the stability of the spectral decimation algorithm.

Indeed, in this case, the RS coincides theoretically with uniform sampling, as $\alpha(s) = p(0) = 1$. This implies that the ratio of accepted values, in each bin, $\alpha_j^{\text{RS}}$ is a constant. However, this does not occur in practice, because of the inherent variance of the Poisson sample. This phenomenon is more evident when either small initial samples are considered or at later stages of the decimation.

For bin $j$ of width $\delta$, the mean Poisson probability is 
\begin{equation}
    \pi_j^{\text{P}} =\frac{1}{\delta}\int_{j\delta}^{(j+1)\delta} \dd s\; e^{-s} =  e^{-j\delta}\frac{1-e^{-\delta}}{\delta},
\end{equation}
and the expected number of counts is $\mathbb{E}[d_j] = d_j^{\text{P}} = d\delta \pi_j^{\text{P}}$, with variance $\operatorname{var}[d_j] = d_j^{\text{P}}$. The actual number of counts in bin $j$ follows a Bernoulli distribution: $d_j \sim \operatorname{Bern}(d, \delta \pi_j^{\text{P}})$.

Therefore, the empirical acceptance ratio $\alpha_j = \mathbb{E}[d_j]/d_j$ has variance $\operatorname{var}[\alpha_j] = 1/d_j$, implying that the standard deviation behaves as:
\begin{equation}
    \sigma_\alpha (s) = \frac{e^{s/2}}{\sqrt{d \delta}}.
\end{equation}
As $\alpha(s)$ is a probability and cannot take values larger than $1$, statistical noise becomes stronger as $s$ increases. In particular, if one chooses a threshold $\sigma_\alpha(s) = 1-\epsilon$, one obtains noisy acceptance probabilities starting from  
\begin{equation}\label{e_appendix_variance_resampling}
    s_{\text{noise}}(\epsilon) = \log (d \delta) - 2 \log (1-\epsilon).
\end{equation}
At the value $s_{\text{noise}}(1) = \log (d \delta)$, the acceptance probability has the wildest oscillations, and it is possible that true Poisson gaps are refused only because of statistical noise.

A practical solution to this problem is to perform a normality test between the empirical counts and the expected Poisson counts. As this should only stabilize true Poisson samples, it is independent with the following logic:
\begin{itemize}
    \item One checks if the first bin counts are compatible with the Poisson estimate;
    \begin{itemize}
        \item if the condition is true, one proceeds with the checks for the other bins. If the bin, say $j$, has counts compatible with the Poisson expectation, then the acceptance ratio is set to $\alpha_j \equiv 1$.
        \item if the condition is false, implying that the counts are not compatible with the Poisson estimate, the acceptance ratio for that bin is left untouched.
    \end{itemize}
\end{itemize}

\subsection{Variance of the Output after Resamplings}

Let us consider the application of multiple decimations on the same spectrum of size $d$. Let the expectation value of the output be $d_{\text{out}} = D$, implying that the Poisson fraction is $M = 1 - D/d$. As the decimation algorithm is probabilistic, which is the variance of $d_{\text{out}}$ obtained after resampling the spectrum $R$ times?

When the spectrum has been decimated to the size $D$, the number of extracted Poisson gaps in the first bin has expectation $\mathbb{E}[d_0] = M d \delta_D$, where $\delta_D$ is the optimal bin-width according to the Freedman-Diaconis rule.

The same quantity $d_0$ can be computed by tracking the number of Poisson gaps extracted at each iteration. At iteration $m_*$, the spectrum has been decimated to $D$, thus $D = d (1-f)^{m_*}$ and 
\begin{equation}
    m_* = \frac{\ln(D/d)}{\ln(1-f)}.
\end{equation}
Henceforth $d_0$ will be the sum of the Poisson gaps extracted at every iteration up to $m_*$, in average:
\begin{equation}
    \mathbb{E}[d_0] = \sum_{m=0}^{m_*} d_E(m) \delta(m) = 12^{1/3} \frac{f}{1-(1-f)^{2/3}} \left(d^{2/3}-D^{2/3}\right) \underset{f\to 0}{\sim} \frac{3}{2}12^{1/3}\left(d^{2/3}-D^{2/3}\right).
\end{equation}
It is easily checked that
\begin{equation}
    \operatorname{var}[d_0] = \left(\frac{\dd\mathbb{E}[d_0]}{\dd D}\right)^2 \operatorname{var}[D] = \delta^2_D \operatorname{var}[D].
\end{equation}
As $d_0$ is the sum of Bernoulli variables in the first bin, its variance is $\operatorname{var}[d_0] = \mathbb{E}[d_0]$. This equality leads to 
\begin{equation}
    \operatorname{var}[D] = \frac{M d}{\delta_D} = \left(1 - \frac{D}{d}\right)\frac{d D^{1/3}}{12^{1/3}}.
\end{equation}

\section{Decimation of Statistical Mixtures}\label{a_numerical_checks}

\begin{figure}
    \centering
    \subfloat[]{\includegraphics[width=0.33\linewidth]{
    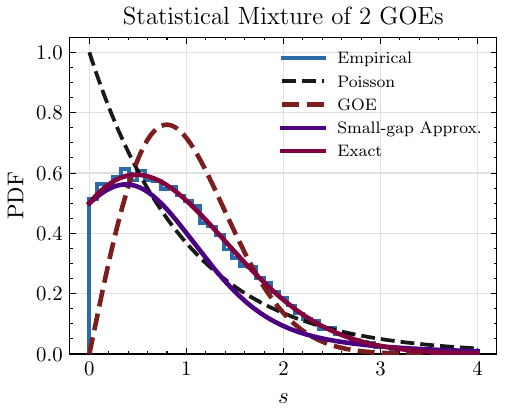
    }}
    \hfill
    \subfloat[]{\includegraphics[width=0.33\linewidth]{
    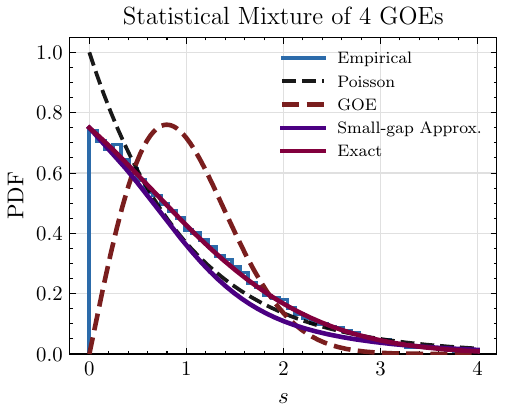
    }}
    \hfill
    \subfloat[]{\includegraphics[width=0.33\linewidth]{
    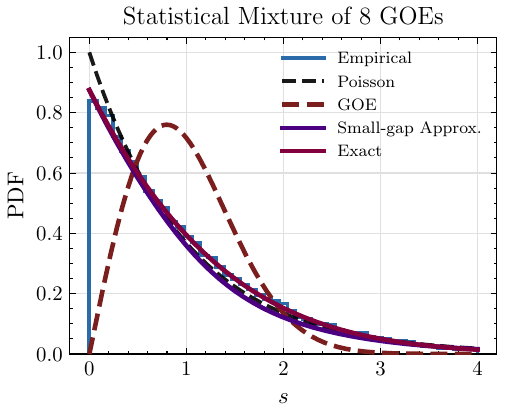
    }}
    \caption{Statistical Mixture of $N$ GOEs of the same size $d_{\text{GOE}} = 10^4$, respectively $N = 2$ (a), $N = 4$ (b) and $N = 8$ (c). At moderately small $s \lesssim 0.5$, the small $s$ approximation, see Section~\ref{s_mixtures}, qualitatively describes the gap distribution, whose exact PDF \cite{Porter2, BerryRobnik, Mehta1} is shown. In particular, it predicts the correct value of the PDF near the origin. As $N$ increasing, the Poisson component of the mixture becomes dominant with respect to the non-Poissonian one, featuring level repulsion, and the approximation becomes better in describing the mixture for larger values of $s$.}
    \label{f_small_approximation}
\end{figure}

This section provides a check of the stability of the decimation algorithm. The setup is studying a generic statistical mixture with $N$ GOE spectra, each of them of size $d_i = 10^4$. Gap statistics for $N=2$, $4$ and $8$ are reported in Fig.~\ref{f_small_approximation} and compared with the small-$s$ approximation of Section~\ref{s_mixtures}.

\paragraph{One-Sample Analysis.}
For a mixture of $N$ GOE spectra of the same dimension $d_{\text{GOE}}$, Eq.~\eqref{e_expectation_dout} predicts that the spectral decimation algorithm output, $d_{\text{out}, N}^{(1)}$ to be compatible with $d_{\text{GOE}}$. However, $d_{\text{out}, N}^{(1)}$ is a random variable, as can be easily checked by repeating the decimation on the same spectrum multiple times (resample). 

For one sample, the Poisson fraction, whose expectation is $\mathbb{E}[M] = 1-1/N$, is itself a random variable, quantified by the number of counts in the first bin: $d_0 = M \delta d$. As $\operatorname{var}[d_0] = d_0$, one obtains that $\operatorname{var}[M] = \frac{M}{\delta d}$. Hence 
\begin{equation}\label{e_appendix_variance_multiple}
    \operatorname{var}[d_{\text{out}, N}^{(1)}] = \left(1 - \frac{1}{N}\right)\frac{d}{\delta} = 12^{-1/3} \left(1 - \frac{1}{N}\right) \left(N d_{\text{GOE}}\right)^{4/3}.
\end{equation} 
For large enough $N$, the standard deviation grows as $N^{1/3}$, implying that $d_{\text{out}, N}^{(1)}$ can naively differ very much from $d_{\text{GOE}}$, while still being compatible.

\begin{figure}
    \centering
    \subfloat[]{\includegraphics[width=0.48\linewidth]{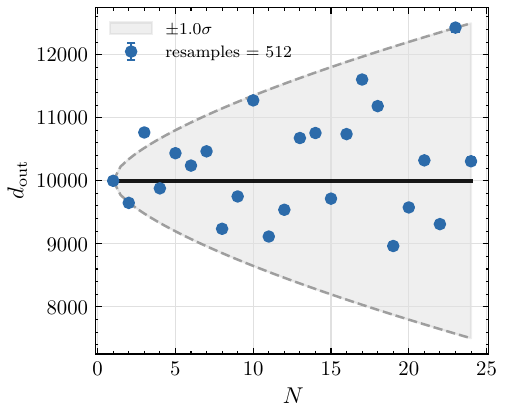}}
    \hfill
    \subfloat[]{\includegraphics[width=0.48\linewidth]{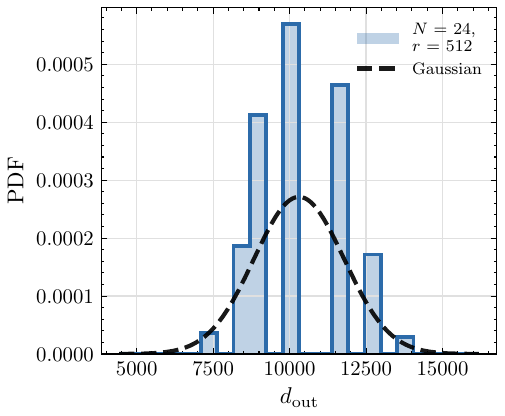}}
    \caption{Decimation of a single realization of a statistical mixture, with $d_{\text{GOE}} = 1$. (a) After 512 resamples for each $N \leq 24$, $d_{\text{out}}$ is plotted against the expected value $d_{\text{GOE}}$. Almost all data points fall within one standard deviation from the mean. (b) Distribution of $d_{\text{out}}$ for $N=24$, resampled $r = 512$ times. Overall, the distribution follows a Gaussian with variance predicted from the spectral decimation, Eq.~\eqref{e_appendix_variance_resampling}. Because of the discrete extraction fraction values, $d_{\text{out}}$ can only assume discrete values, thus the peaks in the PDF.}
    \label{f_one_sample_num_check}
\end{figure}

This prediction can be easily checked with a numerical experiment, see Fig.~\ref{f_one_sample_num_check}. The knowledge of one sample, even if resampled many times, is not sufficient to estimate $d_{\text{GOE}}$ with arbitrary accuracy. To this aim, the analysis of multiple samples is needed.

\paragraph{Analysis of Multiple Samples}

The analysis of $t$ samples of mixtures with the same $N$ can be performed by generating $(tN)$ GOE spectra of dimension $d_{\text{GOE}}$. It is also possible to diagonalize a sufficiently large number of GOE matrices -- in our case 512, -- and construct the mixture by superposing $N$ randomly chosen GOE spectra from the original pool. Because of computational efficiency, we resort to the last approach to generate the $t$ samples. 

The goal of this analysis is to assess whether the decimation can predict $d_{\text{GOE}}$ accurately, up to the finite-size artifacts which are intrinsic to the statistics. These errors can be significantly reduced by pooling the $t$ samples together since we are interested only in the average $d_{\text{out}}$. 

\begin{figure}
    \centering
    \subfloat[]{\includegraphics[width=0.48\linewidth]{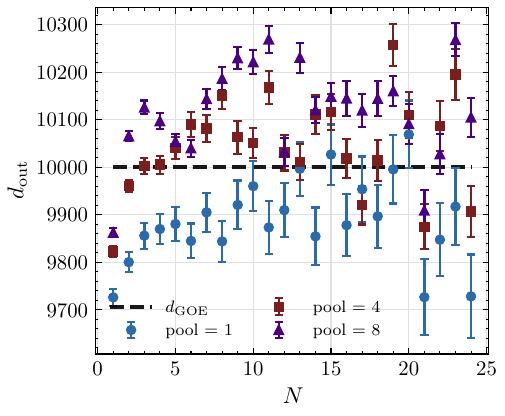}}
    \hfill
    \subfloat[]{\includegraphics[width=0.48\linewidth]{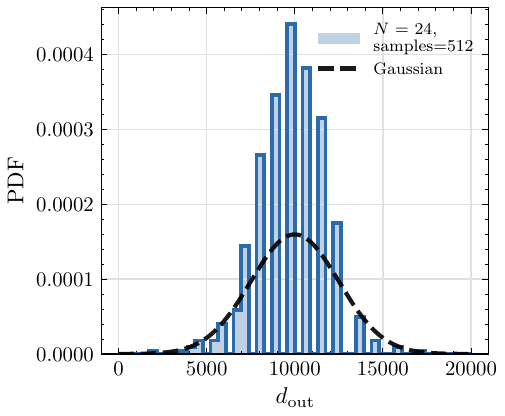}}
    \caption{Result of the decimation procedure on statistical mixtures generated by $N$ identical GOEs of size $d_{\text{GOE}} = 10^4$. (a) Decimation algorithm output, $d_{\text{out}}$ as function of the number $N$. Each data point is obtained by decimating 512 mixtures, individually or by pooling them in groups of $p$. For each pooling $p$, data are contained within a window of $3\%$ of the true value $d_{\text{GOE}}$. (b) Focusing on $N=24$ and $p=1$, we check the distribution of the decimation outputs and observe that they are distributed according to a Gaussian with mean $d_{\text{GOE}}$ and variance Eq.~\eqref{e_appendix_variance_multiple}. As both the initial size and extraction fraction are fixed, the possible $d_{\text{out}}$ are discrete, hence the reason for the peaks.}
    \label{f_num_check_GOEs}
\end{figure}

Fig.~\ref{f_num_check_GOEs} shows that $d_{\text{GOE}}$ can be accurately estimated from the decimation procedure and that the pooling technique is equivalent to averaging the $d_{\text{out}}$'s of individual samples. For each $N$, the outcomes of the decimation follow a Gaussian distribution with a mean of $d_{\text{GOE}}$ and variance given by Eq.~\eqref{e_appendix_variance_multiple}.

\paragraph{Dynamical Signatures of the CSS}

We conclude this Appendix by considering the behavior of dynamical probes of quantum ergodicity, such as the spectral form factor (SFF), defined as
\begin{equation}
    K(t) = \Big\langle \sum_{m,n} e^{it(E_m-E_n)} \Big\rangle. 
\end{equation}
The SFF is a measure of spectral rigidity, also known as two-level form factor \cite{Mehta1}, which depends not only on short-range correlations, and also on additional properties of the Hamiltonian, including long-range spectral correlations and global density-of-states effects. 

\begin{figure}
    \centering
    {\includegraphics[width=0.48\linewidth]{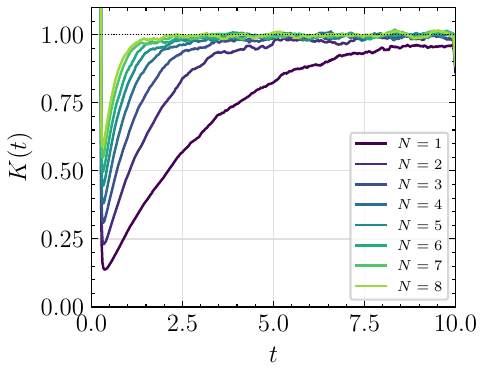}}
    \caption{Spectral Form Factor (SFF) of the Hamiltonians $H_N$, obtained as mixtures of $N$ GOE's of size $10^4$. For each line, $m = 10^3$ eigenvalues are sampled in the bulk of the density of states -- Wigner's semicircle. We observe that }
    \label{f_SFF}
\end{figure}

We consider Hamiltonians $H_N$ whose spectra are obtained as statistical mixtures of $N$ independent GOE spectra, each of size $10^4$. Since all ensembles share the same semicircular density of states, we focused on $m= 10^3$ consecutive unfolded levels in the bulk and computed the corresponding SFF for different values of $N$. The results were further averaged over disorder realizations and locally smoothed in time in order to resolve the ramp structure. Its plot, for $ N \leq 8$ is given in Fig.~\ref{f_SFF}. 

The effective size of the CSS in this construction is $d(N) = m/N$. Fig.~\ref{f_SFF} shows a clear and systematic dependence of the SFF on the CSS size. As $d(N)$ decreases from its maximal value at $N = 1$ --- corresponding to a pure GOE spectrum --- two characteristic effects emerge:
\begin{itemize}
\item the ramp of the SFF becomes progressively steeper;
\item the depth of the correlation hole is strongly suppressed.
\end{itemize}

Together, these effects continuously drive the SFF from the GOE behavior toward the Poisson limit $K_{\text{Poisson}}(t) = 1$, which is recovered as $N\to\infty$, namely when the spectrum becomes completely uncorrelated.

\printbibliography

@misc{statisticalsignatures,
      title="{Statistical Signatures of Integrable and Non-Integrable Quantum Hamiltonians}", 
      author={Feng He and Arthur Hutsalyuk and Giuseppe Mussardo and Andrea Stampiggi},
      year={2025},
      eprint={2510.02440},
      archivePrefix={arXiv},
      primaryClass={cond-mat.stat-mech}
}

@article{HilbertSpaceFragmentationMotrunich,
  title = "{Hilbert Space Fragmentation and Commutant Algebras}",
  author = {Moudgalya, Sanjay and Motrunich, Olexei I.},
  journal = {Phys. Rev. X},
  volume = {12},
  issue = {1},
  pages = {011050},
  numpages = {44},
  year = {2022},
  month = {Mar},
  publisher = {American Physical Society},
  doi = {10.1103/PhysRevX.12.011050}
}

@article{FreedmanDiaconis,
  title = "{On the Histogram as a Density Estimator:$L_2$ Theory}",
  author = {Freedman, David and Diaconis, Persi},
  journal = {Probab. Theory Relat. Fields},
  date = {1981-12},
  volume = {57},
  number = {4},
  pages = {453--476},
  issn = {0044-3719, 1432-2064},
  doi = {10.1007/BF01025868},
  url = {http://link.springer.com/10.1007/BF01025868},
  urldate = {2025-09-03},
  langid = {english}
}

@article{Berry,
author={M.V. Berry and M. Tabor},
  title = {Level Clustering in the Regular Spectrum},
  date = {1977-09-15},
  year={1977},
  journaltitle = {Proc. R. Soc. Lond. A},
  volume = {356},
  number = {1686},
  pages = {375--394},
  issn = {0080-4630},
  doi = {10.1098/rspa.1977.0140},
  url = {https://royalsocietypublishing.org/doi/10.1098/rspa.1977.0140},
  urldate = {2025-09-03},
  langid = {english}
}

@article{prosen,
  title = "{Many-body localization in the Heisenberg $XXZ$ magnet in a random field}",
  author = {\v{Z}nidari\v{c}, Marko and Prosen, Toma\v{z} Prelov\v{s}ek, Peter},
  journal = {Phys. Rev. B},
  volume = {77},
  issue = {6},
  pages = {064426},
  numpages = {5},
  year = {2008},
  month = {Feb},
  publisher = {American Physical Society},
  doi = {10.1103/PhysRevB.77.064426},
  url = {https://link.aps.org/doi/10.1103/PhysRevB.77.064426}
}

@article{huse,
  title = "{Many-body localization phase transition}",
  author = {Pal, Arijeet and Huse, David A.},
  journal = {Phys. Rev. B},
  volume = {82},
  issue = {17},
  pages = {174411},
  numpages = {7},
  year = {2010},
  month = {Nov},
  publisher = {American Physical Society},
  doi = {10.1103/PhysRevB.82.174411},
  url = {https://link.aps.org/doi/10.1103/PhysRevB.82.174411}
}

@article{alet,
  title = "{Many-body localization edge in the random-field Heisenberg chain}",
  author = {Luitz, David J. and Laflorencie, Nicolas and Alet, Fabien},
  journal = {Phys. Rev. B},
  volume = {91},
  issue = {8},
  pages = {081103},
  numpages = {5},
  year = {2015},
  month = {Feb},
  publisher = {American Physical Society},
  doi = {10.1103/PhysRevB.91.081103},
  url = {https://link.aps.org/doi/10.1103/PhysRevB.91.081103}
}

@article{ros1,
title = "{Integrals of motion in the many-body localized phase}",
journal = {Nucl. Phys. B},
volume = {891},
pages = {420-465},
year = {2015},
issn = {0550-3213},
doi = {https://doi.org/10.1016/j.nuclphysb.2014.12.014},
url = {https://www.sciencedirect.com/science/article/pii/S0550321314003836},
author = {V. Ros and M. Müller and A. Scardicchio}
}

@article{ros2,
author = {Imbrie, John Z. and Ros, Valentina and Scardicchio, Antonello},
title = "{Local integrals of motion in many-body localized systems}",
journal = {Annalen der Physik},
volume = {529},
number = {7},
pages = {1600278},
doi = {https://doi.org/10.1002/andp.201600278},
year = {2017}
}

@article{randomcouplingXXXscardicchioabanin,
  title = "{Non-Abelian Symmetries and Disorder: A Broad Nonergodic Regime and Anomalous Thermalization}",
  author = {Protopopov, Ivan V. and Panda, Rajat K. and Parolini, Tommaso and Scardicchio, Antonello and Demler, Eugene and Abanin, Dmitry A.},
  journal = {Phys. Rev. X},
  volume = {10},
  issue = {1},
  pages = {011025},
  numpages = {26},
  year = {2020},
  month = {Feb},
  publisher = {American Physical Society},
  doi = {10.1103/PhysRevX.10.011025},
  url = {https://link.aps.org/doi/10.1103/PhysRevX.10.011025}
}

@article{Density-1,
  title = "{Density matrix formulation for quantum renormalization groups}",
  author = {White, Steven R.},
  journal = {Phys. Rev. Lett.},
  volume = {69},
  issue = {19},
  pages = {2863--2866},
  numpages = {0},
  year = {1992},
  month = {Nov},
  publisher = {American Physical Society},
  doi = {10.1103/PhysRevLett.69.2863},
  url = {https://link.aps.org/doi/10.1103/PhysRevLett.69.2863}
}

@article{Density-2,
  title = "{Density-matrix algorithms for quantum renormalization groups}",
  author = {White, Steven R.},
  journal = {Phys. Rev. B},
  volume = {48},
  issue = {14},
  pages = {10345--10356},
  numpages = {0},
  year = {1993},
  month = {Oct},
  publisher = {American Physical Society},
  doi = {10.1103/PhysRevB.48.10345},
  url = {https://link.aps.org/doi/10.1103/PhysRevB.48.10345}
}

@article{schollwock2011density,
  title={The density-matrix renormalization group in the age of matrix product states},
  author={Schollw{\"o}ck, Ulrich},
  journal={Annals of physics},
  volume={326},
  number={1},
  pages={96--192},
  year={2011},
  publisher={Elsevier},
  doi = {https://doi.org/10.1016/j.aop.2010.09.012},
  url = {https://www.sciencedirect.com/science/article/abs/pii/S0003491610001752}
}

@Article{Tensor-network,
	title={{The ITensor Software Library for Tensor Network Calculations}},
	author={Matthew Fishman and Steven R. White and E. Miles Stoudenmire},
	journal={SciPost Phys. Codebases},
	pages={4},
	year={2022},
	publisher={SciPost},
	doi={10.21468/SciPostPhysCodeb.4},
	url={https://scipost.org/10.21468/SciPostPhysCodeb.4},
}

@article{Renorm-1,
  title = "{Functional renormalization group for nonequilibrium quantum many-body problems}",
  author = {Gezzi, R. and Pruschke, Th. and Meden, V.},
  journal = {Phys. Rev. B},
  volume = {75},
  issue = {4},
  pages = {045324},
  numpages = {14},
  year = {2007},
  month = {Jan},
  publisher = {American Physical Society},
  doi = {10.1103/PhysRevB.75.045324},
  url = {https://link.aps.org/doi/10.1103/PhysRevB.75.045324}
}

@article{Renorm-2,
  title = "{Renormalization algorithms for Quantum-Many Body Systems in two and higher dimensions}",
  journal = {},
  author = {Verstraete, F. and Cirac, J. I.},
  year={2004},
  eprint={cond-mat/0407066},
  archivePrefix={arXiv}
}

@article{neural5,
author = {Giuseppe Carleo  and Matthias Troyer},
title = "{Solving the quantum many-body problem with artificial neural networks}",
journal = {Science},
volume = {355},
number = {6325},
pages = {602-606},
year = {2017},
doi = {10.1126/science.aag2302},
URL = {https://www.science.org/doi/abs/10.1126/science.aag2302},
eprint = {https://www.science.org/doi/pdf/10.1126/science.aag2302}
}

@article{neural4,
title = "{Approximation capabilities of multilayer feedforward networks}",
journal = {Neural Networks},
volume = {4},
number = {2},
pages = {251-257},
year = {1991},
issn = {0893-6080},
doi = {https://doi.org/10.1016/0893-6080(91)90009-T},
url = {https://www.sciencedirect.com/science/article/pii/089360809190009T},
author = {Kurt Hornik}
}

@article{neural1,
  title = "{Neural network wave functions and the sign problem}",
  author = {Szab\'o, Attila and Castelnovo, Claudio},
  journal = {Phys. Rev. Res.},
  volume = {2},
  issue = {3},
  pages = {033075},
  numpages = {12},
  year = {2020},
  month = {Jul},
  publisher = {American Physical Society},
  doi = {10.1103/PhysRevResearch.2.033075},
  url = {https://link.aps.org/doi/10.1103/PhysRevResearch.2.033075}
}

@article{neural2,
  title = "{Dirac-Type Nodal Spin Liquid Revealed by Refined Quantum Many-Body Solver Using Neural-Network Wave Function, Correlation Ratio, and Level Spectroscopy}",
  author = {Nomura, Yusuke and Imada, Masatoshi},
  journal = {Phys. Rev. X},
  volume = {11},
  issue = {3},
  pages = {031034},
  numpages = {19},
  year = {2021},
  month = {Aug},
  publisher = {American Physical Society},
  doi = {10.1103/PhysRevX.11.031034},
  url = {https://link.aps.org/doi/10.1103/PhysRevX.11.031034}
}

@article{neural3,
  title = "{Quantum Entanglement in Neural Network States}",
  author = {Deng, Dong-Ling and Li, Xiaopeng and Das Sarma, S.},
  journal = {Phys. Rev. X},
  volume = {7},
  issue = {2},
  pages = {021021},
  numpages = {17},
  year = {2017},
  month = {May},
  publisher = {American Physical Society},
  doi = {10.1103/PhysRevX.7.021021},
  url = {https://link.aps.org/doi/10.1103/PhysRevX.7.021021}
}

@article{neural-I,
  title = "{Efficient representation of quantum many-body states with deep neural networks}",
  author = {Gao, Xun and Duan, Lu-Ming},
  journal = {Nature Comm.},
  volume = {8},
  issue = {1},
  pages = {662},
  year = {2017},
  month = {09},
  doi = {10.1038/s41467-017-00705-2},
  url = {https://doi.org/10.1038/s41467-017-00705-2}
}

@article{neural-II,
  title = "{Constructing exact representations of quantum many-body systems with deep neural networks}",
  author = {Carleo, Giuseppe and Nomura, Yusuke and Imada, Masatoshi},
  journal = {Nature Comm.},
  volume = {9},
  issue = {1},
  pages = {5322},
  year = {2018},
  month = {12},
  doi = {10.1038/s41467-018-07520-3},
  url = {https://doi.org/10.1038/s41467-018-07520-3}
}

@book{Porter,
  title = {Statistical {{Theory}} of {{Spectra}}: {{Fluctuations}}},
  author = {Porter, Charls E.},
  date = {1965},
  publisher = {Elsevier Science \& Technology Books},
  isbn = {978-0-12-562356-8}
}

@article{Porter2,
  title = {"{{Repulsion}} of {{Energy Levels}}" in {{Complex Atomic Spectra}}},
  author = {Rosenzweig, Norbert and Porter, Charles E.},
  date = {1960-12-01},
  journal = {Phys. Rev.},
  volume = {120},
  number = {5},
  pages = {1698--1714},
  issn = {0031-899X},
  doi = {10.1103/PhysRev.120.1698},
  langid = {english}
}

@article{DysonI,
  title = {Statistical {{Theory}} of the {{Energy Levels}} of {{Complex Systems}}. {{I}}},
  author = {Dyson, Freeman J.},
  year = {1962},
  journal = {J. Math. Phys.},
  volume = {3},
  number = {1},
  pages = {140},
  doi = {10.1063/1.1703773}
}

@article{DysonII,
  title = "{Statistical Theory of the Energy Levels of Complex Systems. II}",
  author = {Dyson, Freeman J.},
  journal = {J. Math. Phys.},
  volume = {3},
  number = {1},
  pages = {157},
  year = {1962},
  doi = {10.1063/1.1703774}
}

@article{Vernier,
  title = {Probing {{Symmetries}} of {{Quantum Many-Body Systems}} through {{Gap Ratio Statistics}}},
  author = {Giraud, Olivier and Mac\'{e}, Nicolas and Vernier, \'{E}ric and Alet, Fabien},
  date = {2022-01-10},
  journal = {Phys. Rev. X},
  volume = {12},
  number = {1},
  pages = {011006},
  issn = {2160-3308},
  doi = {10.1103/PhysRevX.12.011006},
  urldate = {2025-09-03},
  langid = {english}
}

@article{Deutsch1,
  title = "{Quantum statistical mechanics in a closed system}",
  author = {Deutsch, J. M.},
  journal = {Phys. Rev. A},
  volume = {43},
  issue = {4},
  pages = {2046--2049},
  numpages = {0},
  year = {1991},
  month = {Feb},
  publisher = {American Physical Society},
  doi = {10.1103/PhysRevA.43.2046},
  url = {https://link.aps.org/doi/10.1103/PhysRevA.43.2046}
}

@article{Deutsch2,
doi = {10.1088/1361-6633/aac9f1},
year = {2018},
month = {jul},
publisher = {IOP Publishing},
volume = {81},
number = {8},
pages = {082001},
author = {Deutsch, Joshua M},
title = {Eigenstate thermalization hypothesis},
journal = {Rep. Prog. Phys.}
}

@article{Srednicki,
  title = "{Chaos and quantum thermalization}",
  author = {Srednicki, Mark},
  journal = {Phys. Rev. E},
  volume = {50},
  issue = {2},
  pages = {888--901},
  numpages = {0},
  year = {1994},
  month = {Aug},
  publisher = {American Physical Society},
  doi = {10.1103/PhysRevE.50.888},
  url = {https://link.aps.org/doi/10.1103/PhysRevE.50.888}
}

@article{Rigol1,
  title = "{Breakdown of Thermalization in Finite One-Dimensional Systems}",
  author = {Rigol, Marcos},
  journal = {Phys. Rev. Lett.},
  volume = {103},
  issue = {10},
  pages = {100403},
  numpages = {4},
  year = {2009},
  month = {Sep},
  publisher = {American Physical Society},
  doi = {10.1103/PhysRevLett.103.100403},
  url = {https://link.aps.org/doi/10.1103/PhysRevLett.103.100403}
}

@article{Rigol2,
  title = "{Onset of quantum chaos in one-dimensional bosonic and fermionic systems and its relation to thermalization}",
  author = {Santos, Lea F. and Rigol, Marcos},
  journal = {Phys. Rev. E},
  volume = {81},
  issue = {3},
  pages = {036206},
  numpages = {10},
  year = {2010},
  month = {Mar},
  publisher = {American Physical Society},
  doi = {10.1103/PhysRevE.81.036206},
  url = {https://link.aps.org/doi/10.1103/PhysRevE.81.036206}
}

@article{DeLuca,
doi = {10.1209/0295-5075/101/37003},
url = {https://doi.org/10.1209/0295-5075/101/37003},
year = {2013},
month = {feb},
publisher = {EDP Sciences, IOP Publishing and Società Italiana di Fisica},
volume = {101},
number = {3},
pages = {37003},
author = {De Luca, A. and Scardicchio, A.},
title = "{Ergodicity breaking in a model showing many-body localization}",
journal = {Europhys. Lett.}
}

@article{Vidmar1,
  title = "{Quantum chaos challenges many-body localization}",
  author = {\v{S}untajs, Jan and Bon\v{c}a, Janez and Prosen, Toma\v{z}  and Vidmar, Lev},
  journal = {Phys. Rev. E},
  volume = {102},
  issue = {6},
  pages = {062144},
  numpages = {12},
  year = {2020},
  month = {Dec},
  publisher = {American Physical Society},
  doi = {10.1103/PhysRevE.102.062144}
}

@article{Scardicchio3,
  title = "{The many-body localized phase of the quantum random energy model}",
  author = {Baldwin, C. L. and Laumann, C. R. and Pal, A. and Scardicchio, A.},
  journal = {Phys. Rev. B},
  volume = {93},
  issue = {2},
  pages = {024202},
  numpages = {15},
  year = {2016},
  month = {Jan},
  publisher = {American Physical Society},
  doi = {10.1103/PhysRevB.93.024202}
}

@article{De_2022,
doi = {10.1088/1751-8121/ac39cd},
year = {2021},
month = {dec},
publisher = {IOP Publishing},
volume = {55},
number = {1},
pages = {014001},
author = {De, Bitan and Sierant, Piotr and Zakrzewski, Jakub},
title = "{On intermediate statistics across many-body localization transition}",
journal = {J. Phys. A: Math. Theor.}
}

@article{SierantDelande,
  title = "{Many-body localization due to random interactions}",
  author = {Sierant, Piotr and Delande, Dominique and Zakrzewski, Jakub},
  journal = {Phys. Rev. A},
  volume = {95},
  issue = {2},
  pages = {021601},
  numpages = {5},
  year = {2017},
  month = {Feb},
  publisher = {American Physical Society},
  doi = {10.1103/PhysRevA.95.021601}
}

@article{Pollmann1,
  title = "{Dynamics of strongly interacting systems: From Fock-space fragmentation to many-body localization}",
  author = {De Tomasi, Giuseppe and Hetterich, Daniel and Sala, Pablo and Pollmann, Frank},
  journal = {Phys. Rev. B},
  volume = {100},
  issue = {21},
  pages = {214313},
  numpages = {15},
  year = {2019},
  month = {Dec},
  publisher = {American Physical Society},
  doi = {10.1103/PhysRevB.100.214313}
}

@article{Pollmann2,
  title = "{Statistical localization: From strong fragmentation to strong edge modes}",
  author = {Rakovszky, Tibor and Sala, Pablo and Verresen, Ruben and Knap, Michael and Pollmann, Frank},
  journal = {Phys. Rev. B},
  volume = {101},
  issue = {12},
  pages = {125126},
  numpages = {23},
  year = {2020},
  month = {Mar},
  publisher = {American Physical Society},
  doi = {10.1103/PhysRevB.101.125126}
}

@article{Pollmann3,
  title = {Hilbert space fragmentation in open quantum systems},
  author = {Li, Yahui and Sala, Pablo and Pollmann, Frank},
  journal = {Phys. Rev. Res.},
  volume = {5},
  issue = {4},
  pages = {043239},
  numpages = {17},
  year = {2023},
  month = {Dec},
  publisher = {American Physical Society},
  doi = {10.1103/PhysRevResearch.5.043239}
}

@article{Q-East,
  title = "{Aspects of Hilbert space fragmentation in the quantum East model: Fragmentation, subspace-restricted quantum scars, and effects of density-density interactions}",
  author = {Ganguli, Maitri and Aditya, Sreemayee and Sen, Diptiman},
  journal = {Phys. Rev. B},
  volume = {111},
  issue = {4},
  pages = {045411},
  numpages = {25},
  year = {2025},
  month = {Jan},
  publisher = {American Physical Society},
  doi = {10.1103/PhysRevB.111.045411}
}

@article{fragment1,
  title = "{Subspace-restricted thermalization in a correlated-hopping model with strong Hilbert space fragmentation characterized by irreducible strings}",
  author = {Aditya, Sreemayee and Dhar, Deepak and Sen, Diptiman},
  journal = {Phys. Rev. B},
  volume = {110},
  issue = {4},
  pages = {045418},
  numpages = {19},
  year = {2024},
  month = {Jul},
  publisher = {American Physical Society},
  doi = {10.1103/PhysRevB.110.045418}
}

@article{fragment2,
  title = "{Observation of Quantum Thermalization Restricted to Hilbert Space Fragments and ${\mathbb{Z}}_{2k}$ Scars}",
  author = {Zhao, Luheng and Datla, Prithvi Raj and Tian, Weikun and Aliyu, Mohammad Mujahid and Loh, Huanqian},
  journal = {Phys. Rev. X},
  volume = {15},
  issue = {1},
  pages = {011035},
  numpages = {18},
  year = {2025},
  month = {Feb},
  publisher = {American Physical Society},
  doi = {10.1103/PhysRevX.15.011035}
}

@article{fragment3,
  title = "{Glassy Word Problems: Ultraslow Relaxation, Hilbert Space Jamming, and Computational Complexity}",
  author = {Balasubramanian, Shankar and Gopalakrishnan, Sarang and Khudorozhkov, Alexey and Lake, Ethan},
  journal = {Phys. Rev. X},
  volume = {14},
  issue = {2},
  pages = {021034},
  numpages = {47},
  year = {2024},
  month = {May},
  publisher = {American Physical Society},
  doi = {10.1103/PhysRevX.14.021034}
}

@article{fragment4,
  title = "{Minimal Hubbard Models of Maximal Hilbert Space Fragmentation}",
  author = {Kwan, Yves H. and Wilhelm, Patrick H. and Biswas, Sounak and Parameswaran, S. A.},
  journal = {Phys. Rev. Lett.},
  volume = {134},
  issue = {1},
  pages = {010411},
  numpages = {8},
  year = {2025},
  month = {Jan},
  publisher = {American Physical Society},
  doi = {10.1103/PhysRevLett.134.010411}
}

@article{Ising-frag1,
  title = "{Emergence of Hilbert Space Fragmentation in Ising Models with a Weak Transverse Field}",
  author = {Yoshinaga, Atsuki and Hakoshima, Hideaki and Imoto, Takashi and Matsuzaki, Yuichiro and Hamazaki, Ryusuke},
  journal = {Phys. Rev. Lett.},
  volume = {129},
  issue = {9},
  pages = {090602},
  numpages = {8},
  year = {2022},
  month = {Aug},
  publisher = {American Physical Society},
  doi = {10.1103/PhysRevLett.129.090602}
}

@article{Ising-frag2,
  title = "{Hilbert space shattering and dynamical freezing in the quantum Ising model}",
  author = {Hart, Oliver and Nandkishore, Rahul},
  journal = {Phys. Rev. B},
  volume = {106},
  issue = {21},
  pages = {214426},
  numpages = {15},
  year = {2022},
  month = {Dec},
  publisher = {American Physical Society},
  doi = {10.1103/PhysRevB.106.214426}
}

@article{Moudgalya_2022,
doi = {10.1088/1361-6633/ac73a0},
url = {https://doi.org/10.1088/1361-6633/ac73a0},
year = {2022},
month = {jul},
publisher = {IOP Publishing},
volume = {85},
number = {8},
pages = {086501},
author = {Moudgalya, Sanjay and Bernevig, B Andrei and Regnault, Nicolas},
title = "{Quantum many-body scars and Hilbert space fragmentation: a review of exact results}",
journal = {Rep. Prog. Phys.}
}

@misc{caha2018pairflipmodel,
      title={The pair-flip model: a very entangled translationally invariant spin chain}, 
      author={Libor Caha and Daniel Nagaj},
      year={2018},
      eprint={1805.07168},
      archivePrefix={arXiv},
      primaryClass={quant-ph},
      url={https://arxiv.org/abs/1805.07168}, 
}

@article{bracci2025,
  title = "{Renormalization group analysis of the many-body localization transition in the random-field {{XXZ}} chain}",
  author = {Niedda, Jacopo and Testasecca, Giacomo Bracci and Magnifico, Giuseppe and Balducci, Federico and Vanoni, Carlo and Scardicchio, Antonello},
  journal = {Phys. Rev. B},
  volume = {112},
  issue = {14},
  pages = {144201},
  numpages = {14},
  year = {2025},
  month = {Oct},
  publisher = {American Physical Society},
  doi = {10.1103/gcwf-jdlr},
  url = {https://link.aps.org/doi/10.1103/gcwf-jdlr}
}

@misc{stampiggi_spectral_decimation_2026,
  author       = {Andrea Stampiggi},
  title        = {spectral\_decimation},
  year         = {2026},
  publisher    = {Zenodo},
  doi          = {10.5281/zenodo.20349048},
  url          = {https://doi.org/10.5281/zenodo.20349048}
}

@article{BerryRobnik,
  title = "{Semiclassical Level Spacings When Regular and Chaotic Orbits Coexist}",
  author = {Berry, M V and Robnik, M},
  date = {1984-08-21},
  journaltitle = {J. Phys. A: Math. Gen.},
  volume = {17},
  number = {12},
  pages = {2413--2421},
  issn = {0305-4470, 1361-6447},
  doi = {10.1088/0305-4470/17/12/013},
  url = {https://iopscience.iop.org/article/10.1088/0305-4470/17/12/013},
  urldate = {2025-09-03}
}

@book{Mehta1,
  title = "{Random Matrices}",
  author = {Mehta, M. L.},
  date = {2004},
  series = {Pure and Applied Mathematics},
  edition = {3rd ed},
  number = {v. 142},
  publisher = {Elsevier/Academic Press},
  location = {Amsterdam San Diego, CA},
  isbn = {978-0-12-088409-4},
  langid = {english}
}

@article{Brody,
  title = "{Random-Matrix Physics: Spectrum and Strength Fluctuations}",
  shorttitle = {Random-Matrix Physics},
  author = {Brody, T. A. and Flores, J. and French, J. B. and Mello, P. A. and Pandey, A. and Wong, S. S. M.},
  date = {1981-07-01},
  journaltitle = {Rev. Mod. Phys.},
  volume = {53},
  number = {3},
  pages = {385--479},
  issn = {0034-6861},
  doi = {10.1103/RevModPhys.53.385},
  url = {https://link.aps.org/doi/10.1103/RevModPhys.53.385},
  urldate = {2025-09-03},
  langid = {english}
}

@article{CalabreseEsslerMussardo2016,
  title        = "{Quantum Integrability in Out of Equilibrium Systems}",
  editor       = {Calabrese, Pasquale and Essler, Fabian H. L. and Mussardo, Giuseppe},
  journal      = {J. Stat. Mech.: Theor. Exp.},
  volume       = {2016},
  number       = {06},
  pages        = {064001},
  year         = {2016},
  doi          = {10.1088/1742-5468/2016/06/064001}
}

@article{PhysRevX.13.011041,
  title = "{Thermalization of Dilute Impurities in One-Dimensional Spin Chains}",
  author = {Sels, Dries and Polkovnikov, Anatoli},
  journal = {Phys. Rev. X},
  volume = {13},
  issue = {1},
  pages = {011041},
  numpages = {21},
  year = {2023},
  month = {Mar},
  publisher = {American Physical Society},
  doi = {10.1103/PhysRevX.13.011041},
  url = {https://link.aps.org/doi/10.1103/PhysRevX.13.011041}
}

@article{PhysRevE.104.054105,
  title = "{Dynamical obstruction to localization in a disordered spin chain}",
  author = {Sels, Dries and Polkovnikov, Anatoli},
  journal = {Phys. Rev. E},
  volume = {104},
  issue = {5},
  pages = {054105},
  numpages = {11},
  year = {2021},
  month = {Nov},
  publisher = {American Physical Society},
  doi = {10.1103/PhysRevE.104.054105},
  url = {https://link.aps.org/doi/10.1103/PhysRevE.104.054105}
}

@article{Panda_2019, title={Can we study the many-body localisation transition?}, volume={128}, ISSN={1286-4854}, url={http://dx.doi.org/10.1209/0295-5075/128/67003}, DOI={10.1209/0295-5075/128/67003}, number={6}, journal={EPL (Europhysics Letters)}, publisher={IOP Publishing}, author={Panda, R. K. and Scardicchio, A. and Schulz, M. and Taylor, S. R. and Žnidarič, M.}, year={2020}, month=feb, pages={67003} }

@book{ran2020tensor,
  title="{Tensor network contractions: methods and applications to quantum many-body systems}",
  author={Ran, Shi-Ju and Tirrito, Emanuele and Peng, Cheng and Chen, Xi and Tagliacozzo, Luca and Su, Gang and Lewenstein, Maciej},
  year={2020},
  doi = {10.1007/978-3-030-34489-4},
  publisher={Springer Nature}
}

@article{orus2019tensor,
  title={Tensor networks for complex quantum systems},
  author={Or{\'u}s, Rom{\'a}n},
  journal={Nature Reviews Physics},
  volume={1},
  number={9},
  pages={538--550},
  year={2019},
  publisher={Nature Publishing Group UK London},
  doi = {https://doi.org/10.1038/s42254-019-0086-7},
  url = {https://www.nature.com/articles/s42254-019-0086-7#citeas}
}

@article{silvi2019tensor,
  title={The Tensor Networks Anthology: Simulation techniques for many-body quantum lattice systems},
  author={Silvi, Pietro and Tschirsich, Ferdinand and Gerster, Matthias and J{\"u}nemann, Johannes and Jaschke, Daniel and Rizzi, Matteo and Montangero, Simone},
  journal={SciPost Physics Lecture Notes},
  pages={008},
  year={2019},
  doi={10.21468/SciPostPhysLectNotes.8},
  url={https://www.scipost.org/SciPostPhysLectNotes.8?acad_field_slug=chemistry}
}

@article{medvidovic2024neural,
  title={Neural-network quantum states for many-body physics},
  author={Medvidovi{\'c}, Matija and Moreno, Javier Robledo},
  journal={The European Physical Journal Plus},
  volume={139},
  number={7},
  pages={1--26},
  year={2024},
  publisher={Springer},
  doi = {https://doi.org/10.1140/epjp/s13360-024-05311-y},
  url = {https://link.springer.com/article/10.1140/epjp/s13360-024-05311-y#article-info}
}

@article{du2025artificial,
  title={Artificial intelligence for representing and characterizing quantum systems},
  author={Du, Yuxuan and Zhu, Yan and Zhang, Yuan-Hang and Hsieh, Min-Hsiu and Rebentrost, Patrick and Gao, Weibo and Wu, Ya-Dong and Eisert, Jens and Chiribella, Giulio and Tao, Dacheng and others},
  journal={arXiv preprint arXiv:2509.04923},
  year={2025},
  url = {https://arxiv.org/abs/2509.04923}
}

@article{PhysRevA.71.012317,
  title = "{Strong many-particle localization and quantum computing with perpetually coupled qubits}",
  author = {Santos, L. F. and Dykman, M. I. and Shapiro, M. and Izrailev, F. M.},
  journal = {Phys. Rev. A},
  volume = {71},
  issue = {1},
  pages = {012317},
  numpages = {18},
  year = {2005},
  month = {Jan},
  publisher = {American Physical Society},
  doi = {10.1103/PhysRevA.71.012317},
  url = {https://link.aps.org/doi/10.1103/PhysRevA.71.012317}
}

@article{PhysRevB.105.224203,
  title = "{Challenges to observation of many-body localization}",
  author = {Sierant, Piotr and Zakrzewski, Jakub},
  journal = {Phys. Rev. B},
  volume = {105},
  issue = {22},
  pages = {224203},
  numpages = {18},
  year = {2022},
  month = {Jun},
  publisher = {American Physical Society},
  doi = {10.1103/PhysRevB.105.224203},
  url = {https://link.aps.org/doi/10.1103/PhysRevB.105.224203}
}

@article{PhysRevB.103.024203,
  title = "{Slow delocalization of particles in many-body localized phases}",
  author = {Kiefer-Emmanouilidis, Maximilian and Unanyan, Razmik and Fleischhauer, Michael and Sirker, Jesko},
  journal = {Phys. Rev. B},
  volume = {103},
  issue = {2},
  pages = {024203},
  numpages = {9},
  year = {2021},
  month = {Jan},
  publisher = {American Physical Society},
  doi = {10.1103/PhysRevB.103.024203},
  url = {https://link.aps.org/doi/10.1103/PhysRevB.103.024203}
}

@article{PhysRevLett.115.187201,
  title = "{Early Breakdown of Area-Law Entanglement at the Many-Body Delocalization Transition}",
  author = {Devakul, Trithep and Singh, Rajiv R. P.},
  journal = {Phys. Rev. Lett.},
  volume = {115},
  issue = {18},
  pages = {187201},
  numpages = {5},
  year = {2015},
  month = {Oct},
  publisher = {American Physical Society},
  doi = {10.1103/PhysRevLett.115.187201},
  url = {https://link.aps.org/doi/10.1103/PhysRevLett.115.187201}
}

@article{PhysRevLett.124.243601,
  title = "{Evidence for Unbounded Growth of the Number Entropy in Many-Body Localized Phases}",
  author = {Kiefer-Emmanouilidis, Maximilian and Unanyan, Razmik and Fleischhauer, Michael and Sirker, Jesko},
  journal = {Phys. Rev. Lett.},
  volume = {124},
  issue = {24},
  pages = {243601},
  numpages = {6},
  year = {2020},
  month = {Jun},
  publisher = {American Physical Society},
  doi = {10.1103/PhysRevLett.124.243601},
  url = {https://link.aps.org/doi/10.1103/PhysRevLett.124.243601}
}

@article{PhysRevB.102.064207,
  title ="{Ergodicity breaking transition in finite disordered spin chains}",
  author = {\v{S}untajs, Jan and Bon\v{c}a, Janez and Prosen, Toma\v{z} and Vidmar, Lev},
  journal = {Phys. Rev. B},
  volume = {102},
  issue = {6},
  pages = {064207},
  numpages = {11},
  year = {2020},
  month = {Aug},
  publisher = {American Physical Society},
  doi = {10.1103/PhysRevB.102.064207},
  url = {https://link.aps.org/doi/10.1103/PhysRevB.102.064207}
}

@article{Pietracaprina_2017,
doi = {10.1088/1742-5468/aa9338},
url = {https://doi.org/10.1088/1742-5468/aa9338},
year = {2017},
month = {nov},
publisher = {IOP Publishing and SISSA},
volume = {2017},
number = {11},
pages = {113102},
author = {Pietracaprina, Francesca and Parisi, Giorgio and Mariano, Angelo and Pascazio, Saverio and Scardicchio, Antonello},
title = "{Entanglement critical length at the many-body localization transition}",
journal = {J. Stat. Mech.: Theory Exp.}
}

@article{10.1098/rsta.2016.0424,
    author = {Mossi, G. and Scardicchio, A.},
    title = "{Ergodic and localized regions in quantum spin glasses on the Bethe lattice}",
    journal = {Philos Trans A Math Phys Eng Sci},
    volume = {375},
    number = {2108},
    pages = {20160424},
    year = {2017},
    month = {10},
    issn = {1364-503X},
    doi = {10.1098/rsta.2016.0424},
    url = {https://doi.org/10.1098/rsta.2016.0424},
}

@article{PhysRevB.107.115132,
  title = "{Stability of many-body localization in Floquet systems}",
  author = {Sierant, Piotr and Lewenstein, Maciej and Scardicchio, Antonello and Zakrzewski, Jakub},
  journal = {Phys. Rev. B},
  volume = {107},
  issue = {11},
  pages = {115132},
  numpages = {16},
  year = {2023},
  month = {Mar},
  publisher = {American Physical Society},
  doi = {10.1103/PhysRevB.107.115132},
  url = {https://link.aps.org/doi/10.1103/PhysRevB.107.115132}
}

@article{PhysRevB.110.184209,
  title = "{Many-body localization crossover is sharper in a quasiperiodic potential}",
  author = {Falc\~ao, Pedro R. Nic\'acio and Aramthottil, Adith Sai and Sierant, Piotr and Zakrzewski, Jakub},
  journal = {Phys. Rev. B},
  volume = {110},
  issue = {18},
  pages = {184209},
  numpages = {14},
  year = {2024},
  month = {Nov},
  publisher = {American Physical Society},
  doi = {10.1103/PhysRevB.110.184209},
  url = {https://link.aps.org/doi/10.1103/PhysRevB.110.184209}
}

@article{PhysRevB.84.094203,
  title = "{Structure of typical states of a disordered Richardson model and many-body localization}",
  author = {Buccheri, F. and De Luca, A. and Scardicchio, A.},
  journal = {Phys. Rev. B},
  volume = {84},
  issue = {9},
  pages = {094203},
  numpages = {9},
  year = {2011},
  month = {Sep},
  publisher = {American Physical Society},
  doi = {10.1103/PhysRevB.84.094203},
  url = {https://link.aps.org/doi/10.1103/PhysRevB.84.094203}
}

@article{PhysRevB.101.214205,
  title = "{Kinetically constrained freezing transition in a dipole-conserving system}",
  author = {Morningstar, Alan and Khemani, Vedika and Huse, David A.},
  journal = {Phys. Rev. B},
  volume = {101},
  issue = {21},
  pages = {214205},
  numpages = {10},
  year = {2020},
  month = {Jun},
  publisher = {American Physical Society},
  doi = {10.1103/PhysRevB.101.214205},
  url = {https://link.aps.org/doi/10.1103/PhysRevB.101.214205}
}

@article{PhysRevLett.128.196601,
  title = "{Fragmentation and Emergent Integrable Transport in the Weakly Tilted Ising Chain}",
  author = {Bastianello, Alvise and Borla, Umberto and Moroz, Sergej},
  journal = {Phys. Rev. Lett.},
  volume = {128},
  issue = {19},
  pages = {196601},
  numpages = {7},
  year = {2022},
  month = {May},
  publisher = {American Physical Society},
  doi = {10.1103/PhysRevLett.128.196601},
  url = {https://link.aps.org/doi/10.1103/PhysRevLett.128.196601}
}

@article{PhysRevB.103.214304,
  title = "{Emergent symmetries and slow quantum dynamics in a Rydberg-atom chain with confinement}",
  author = {Chen, I-Chi and Iadecola, Thomas},
  journal = {Phys. Rev. B},
  volume = {103},
  issue = {21},
  pages = {214304},
  numpages = {13},
  year = {2021},
  month = {Jun},
  publisher = {American Physical Society},
  doi = {10.1103/PhysRevB.103.214304},
  url = {https://link.aps.org/doi/10.1103/PhysRevB.103.214304}
}

@article{PhysRevLett.124.207602,
  title = "{Hilbert-Space Fragmentation from Strict Confinement}",
  author = {Yang, Zhi-Cheng and Liu, Fangli and Gorshkov, Alexey V. and Iadecola, Thomas},
  journal = {Phys. Rev. Lett.},
  volume = {124},
  issue = {20},
  pages = {207602},
  numpages = {6},
  year = {2020},
  month = {May},
  publisher = {American Physical Society},
  doi = {10.1103/PhysRevLett.124.207602},
  url = {https://link.aps.org/doi/10.1103/PhysRevLett.124.207602}
}

@article{PhysRevB.104.155117,
  title = "{Minimal model for Hilbert space fragmentation with local constraints}",
  author = {Mukherjee, Bhaskar and Banerjee, Debasish and Sengupta, K. and Sen, Arnab},
  journal = {Phys. Rev. B},
  volume = {104},
  issue = {15},
  pages = {155117},
  numpages = {20},
  year = {2021},
  month = {Oct},
  publisher = {American Physical Society},
  doi = {10.1103/PhysRevB.104.155117},
  url = {https://link.aps.org/doi/10.1103/PhysRevB.104.155117}
}

@article{PhysRevResearch.3.033201,
  title = "{Constraint-induced breaking and restoration of ergodicity in spin-1 PXP models}",
  author = {Mukherjee, Bhaskar and Cai, Zi and Liu, W. Vincent},
  journal = {Phys. Rev. Res.},
  volume = {3},
  issue = {3},
  pages = {033201},
  numpages = {16},
  year = {2021},
  month = {Aug},
  publisher = {American Physical Society},
  doi = {10.1103/PhysRevResearch.3.033201},
  url = {https://link.aps.org/doi/10.1103/PhysRevResearch.3.033201}
}

@article{PhysRevB.103.235133,
  title = "{Frustration-induced emergent Hilbert space fragmentation}",
  author = {Lee, Kyungmin and Pal, Arijeet and Changlani, Hitesh J.},
  journal = {Phys. Rev. B},
  volume = {103},
  issue = {23},
  pages = {235133},
  numpages = {13},
  year = {2021},
  month = {Jun},
  publisher = {American Physical Society},
  doi = {10.1103/PhysRevB.103.235133},
  url = {https://link.aps.org/doi/10.1103/PhysRevB.103.235133}
}

@Article{10.21468/SciPostPhys.11.4.074,
	title="{Information dynamics in a model with Hilbert space fragmentation}",
	author={Dominik Hahn and Paul A. McClarty and David J. Luitz},
	journal={SciPost Phys.},
	volume={11},
	pages={074},
	year={2021},
	publisher={SciPost},
	doi={10.21468/SciPostPhys.11.4.074},
	url={https://scipost.org/10.21468/SciPostPhys.11.4.074},
}

@article{ZhangZhaoDevakulHuse,
  title = "{Many-body localization phase transition: A simplified strong-randomness approximate renormalization group}",
  author = {Zhang, Liangsheng and Zhao, Bo and Devakul, Trithep and Huse, David A.},
  journal = {Phys. Rev. B},
  volume = {93},
  issue = {22},
  pages = {224201},
  numpages = {7},
  year = {2016},
  month = {Jun},
  publisher = {American Physical Society},
  doi = {10.1103/PhysRevB.93.224201}
}

@article{CorpsMolinaRelano,
	title = "{Signatures of a critical point in the many-body localization transition}",
    author = {\'{A}ngel L. Corps and Rafael A. Molina and Armando Rela\~no},
	journal = {SciPost Phys.},
	volume = {10},
	pages = {107},
	year = {2021},
	publisher = {SciPost},
	doi = {10.21468/SciPostPhys.10.5.107},
	url = {https://scipost.org/10.21468/SciPostPhys.10.5.107}
}

@article{https://doi.org/10.1002/andp.201600360,
author = {Gornyi, I. V. and Mirlin, A. D. and Polyakov, D. G. and Burin, A. L.},
title = "{Spectral diffusion and scaling of many-body delocalization transitions}",
journal = {Annalen der Physik},
volume = {529},
number = {7},
pages = {1600360},
doi = {https://doi.org/10.1002/andp.201600360},
url = {https://onlinelibrary.wiley.com/doi/abs/10.1002/andp.201600360}
}

@article{PhysRevResearch.3.L012019,
  title = "{Universal spectral form factor for many-body localization}",
  author = {Prakash, Abhishodh and Pixley, J. H. and Kulkarni, Manas},
  journal = {Phys. Rev. Res.},
  volume = {3},
  issue = {1},
  pages = {L012019},
  numpages = {6},
  year = {2021},
  month = {Feb},
  publisher = {American Physical Society},
  doi = {10.1103/PhysRevResearch.3.L012019},
  url = {https://link.aps.org/doi/10.1103/PhysRevResearch.3.L012019}
}

@article{PhysRevLett.118.127202,
  title = "{Field Theory Approach to Many-Body Localization}",
  author = {Altland, Alexander and Micklitz, Tobias},
  journal = {Phys. Rev. Lett.},
  volume = {118},
  issue = {12},
  pages = {127202},
  numpages = {5},
  year = {2017},
  month = {Mar},
  publisher = {American Physical Society},
  doi = {10.1103/PhysRevLett.118.127202},
  url = {https://link.aps.org/doi/10.1103/PhysRevLett.118.127202}
}

@article{ModakNag,
  title = "{Many-body localization in a long-range model: Real-space re\-normali\-zation-group study}",
  author = {Modak, Ranjan and Nag, Tanay},
  journal = {Phys. Rev. E},
  volume = {101},
  issue = {5},
  pages = {052108},
  numpages = {12},
  year = {2020},
  month = {May},
  publisher = {American Physical Society},
  doi = {10.1103/PhysRevE.101.052108}
}

@article{VoskAltman2,
   author = "Altman, Ehud and Vosk, Ronen",
   title = "Universal Dynamics and Renormalization in Many-Body-Localized Systems", 
   journal= "Annual Review of Condensed Matter Physics",
   year = "2015",
   volume = "6",
   number = "Volume 6, 2015",
   pages = "383-409",
   doi = "https://doi.org/10.1146/annurev-conmatphys-031214-014701",
   publisher = "Annual Reviews",
   issn = "1947-5462",
   type = "Journal Article"
  }

@article{PotterVasseurParameswaran,
  title = "{Universal Properties of Many-Body Delocalization Transitions}",
  author = {Potter, Andrew C. and Vasseur, Romain and Parameswaran, S. A.},
  journal = {Phys. Rev. X},
  volume = {5},
  issue = {3},
  pages = {031033},
  numpages = {13},
  year = {2015},
  month = {Sep},
  publisher = {American Physical Society},
  doi = {10.1103/PhysRevX.5.031033}
}

@article{PhysRevB.91.085425,
  title = "{Constructing local integrals of motion in the many-body localized phase}",
  author = {Chandran, Anushya and Kim, Isaac H. and Vidal, Guifre and Abanin, Dmitry A.},
  journal = {Phys. Rev. B},
  volume = {91},
  issue = {8},
  pages = {085425},
  numpages = {7},
  year = {2015},
  month = {Feb},
  publisher = {American Physical Society},
  url = {https://link.aps.org/doi/10.1103/PhysRevB.91.085425}
}

@article{o2016explicit,
  title="{Explicit construction of local conserved operators in disordered many-body systems}",
  author={O'Brien, TE and Abanin, DA and Vidal, G and Papic, Z},
  journal={Physical Review B},
  volume={94},
  number={14},
  year={2016},
  publisher={American Physical Society},
  url = {https://journals.aps.org/prb/abstract/10.1103/PhysRevB.94.144208}
}

@article{Agarwal1,
  title = "{Anomalous Diffusion and Griffiths Effects Near the Many-Body Localization Transition}",
  author = {Agarwal, Kartiek and Gopalakrishnan, Sarang and Knap, Michael and M\"uller, Markus and Demler, Eugene},
  journal = {Phys. Rev. Lett.},
  volume = {114},
  issue = {16},
  pages = {160401},
  numpages = {6},
  year = {2015},
  month = {Apr},
  publisher = {American Physical Society},
  doi = {10.1103/PhysRevLett.114.160401},
  url = {https://link.aps.org/doi/10.1103/PhysRevLett.114.160401}
}

@article{DumitrescuVasseurPotter,
  title = "{Scaling Theory of Entanglement at the Many-Body Localization Transition}",
  author = {Dumitrescu, Philipp T. and Vasseur, Romain and Potter, Andrew C.},
  journal = {Phys. Rev. Lett.},
  volume = {119},
  issue = {11},
  pages = {110604},
  numpages = {6},
  year = {2017},
  month = {Sep},
  publisher = {American Physical Society},
  doi = {10.1103/PhysRevLett.119.110604},
  url = {https://link.aps.org/doi/10.1103/PhysRevLett.119.110604}
}

@article{GoremykinaVasseurSerbyn,
  title = "{Analytically Solvable Renormalization Group for the Many-Body Localization Transition}",
  author = {Goremykina, Anna and Vasseur, Romain and Serbyn, Maksym},
  journal = {Phys. Rev. Lett.},
  volume = {122},
  issue = {4},
  pages = {040601},
  numpages = {6},
  year = {2019},
  month = {Jan},
  publisher = {American Physical Society},
  doi = {10.1103/PhysRevLett.122.040601}
}

@misc{MBL-ETH,
      title="{A microscopically motivated renormalization scheme for the MBL/ETH transition}", 
      author={Thiery, Thimoth\'{e}e and M\"{u}ller, Markus and   De Roeck, Wojciech},
      year={2017},
      eprint={1711.09880},
      archivePrefix={arXiv},
      primaryClass={cond-mat}
}

@article{PhysRevB.95.155129,
  title = "{Stability and instability towards delocalization in many-body localization systems}",
  author = {De Roeck, Wojciech and Huveneers, Fran\c{c}ois},
  journal = {Phys. Rev. B},
  volume = {95},
  issue = {15},
  pages = {155129},
  numpages = {14},
  year = {2017},
  month = {Apr},
  publisher = {American Physical Society},
  doi = {10.1103/PhysRevB.95.155129},
  url = {https://link.aps.org/doi/10.1103/PhysRevB.95.155129}
}

@misc{2408.04338,
      title="{Absence of Normal Heat Conduction in Strongly Disordered Interacting Quantum Chains}", 
      author={De Roeck, Wojciech and Giacomin, Lydia and Huveneers, Francois and Prosniak, Oskar},
      year={2024},
      eprint={2408.04338},
      archivePrefix={arXiv},
      primaryClass={math-ph},
      url={https://arxiv.org/abs/2408.04338}, 
}

@article{ThieryMuller,
  title = "{Many-Body Delocalization as a Quantum Avalanche}",
  author = {Thiery, Thimoth\'ee and Huveneers, Fran\c{c}ois and M\"uller, Markus and De Roeck, Wojciech},
  journal = {Phys. Rev. Lett.},
  volume = {121},
  issue = {14},
  pages = {140601},
  numpages = {6},
  year = {2018},
  month = {Oct},
  publisher = {American Physical Society},
  doi = {10.1103/PhysRevLett.121.140601}
}

@article{SierantZakrzewski2,
  title = "{Level statistics across the many-body localization transition}",
  author = {Sierant, Piotr and Zakrzewski, Jakub},
  journal = {Phys. Rev. B},
  volume = {99},
  issue = {10},
  pages = {104205},
  numpages = {18},
  year = {2019},
  month = {Mar},
  publisher = {American Physical Society},
  doi = {10.1103/PhysRevB.99.104205},
  url = {https://link.aps.org/doi/10.1103/PhysRevB.99.104205}
}

@article{HSFPollman,
  title = "{Ergodicity Breaking Arising from Hilbert Space Fragmentation in Dipole-Conserving Hamiltonians}",
  author = {Sala, Pablo and Rakovszky, Tibor and Verresen, Ruben and Knap, Michael and Pollmann, Frank},
  journal = {Phys. Rev. X},
  volume = {10},
  issue = {1},
  pages = {011047},
  numpages = {19},
  year = {2020},
  month = {Feb},
  url={https://journals.aps.org/prx/abstract/10.1103/PhysRevX.10.011047}
  }

@article{Khemani_dipole,
  title = "{Localization from Hilbert space shattering: From theory to physical realizations}",
  author = {Khemani, Vedika and Hermele, Michael and Nandkishore, Rahul},
  journal = {Phys. Rev. B},
  volume = {101},
  issue = {17},
  pages = {174204},
  numpages = {17},
  year = {2020},
  month = {May},
  publisher = {American Physical Society},
  doi = {10.1103/PhysRevB.101.174204},
  url = {https://link.aps.org/doi/10.1103/PhysRevB.101.174204}
}

@inbook{Moudgalya_2021,
   title="{Thermalization and Its Absence within Krylov Subspaces of a Constrained Hamiltonian}",
   ISBN={9789811231711},
   url={http://dx.doi.org/10.1142/9789811231711_0009},
   DOI={10.1142/9789811231711_0009},
   booktitle="{Memorial Volume for Shoucheng Zhang}",
   publisher={WORLD SCIENTIFIC},
   author={Moudgalya, Sanjay and Prem, Abhinav and Nandkishore, Rahul and Regnault, Nicolas and Bernevig, B. Andrei},
   year={2021},
   month=sep, pages={147–209} }

@article{PhysRevX.9.021003,
  title = "{Localization in Fractonic Random Circuits}",
  author = {Pai, Shriya and Pretko, Michael and Nandkishore, Rahul M.},
  journal = {Phys. Rev. X},
  volume = {9},
  issue = {2},
  pages = {021003},
  numpages = {21},
  year = {2019},
  month = {Apr},
  publisher = {American Physical Society},
  doi = {10.1103/PhysRevX.9.021003},
  url = {https://link.aps.org/doi/10.1103/PhysRevX.9.021003}
}

@article{PhysRevB.103.L220304,
  title = "{Hilbert space fragmentation and exact scars of generalized Fredkin spin chains}",
  author = {Langlett, Christopher M. and Xu, Shenglong},
  journal = {Phys. Rev. B},
  volume = {103},
  issue = {22},
  pages = {L220304},
  numpages = {7},
  year = {2021},
  month = {Jun},
  publisher = {American Physical Society},
  doi = {10.1103/PhysRevB.103.L220304},
  url = {https://link.aps.org/doi/10.1103/PhysRevB.103.L220304}
}

@article{PhysRevB.90.174202,
  title = "{Phenomenology of fully many-body-localized systems}",
  author = {Huse, David A. and Nandkishore, Rahul and Oganesyan, Vadim},
  journal = {Phys. Rev. B},
  volume = {90},
  issue = {17},
  pages = {174202},
  numpages = {5},
  year = {2014},
  month = {Nov},
  publisher = {American Physical Society},
  doi = {10.1103/PhysRevB.90.174202},
  url = {https://link.aps.org/doi/10.1103/PhysRevB.90.174202}
}

@article{PhysRevResearch.4.L012003,
  title = {Anomalous hydrodynamics in a class of scarred frustration-free Hamiltonians},
  author = {Richter, Jonas and Pal, Arijeet},
  journal = {Phys. Rev. Res.},
  volume = {4},
  issue = {1},
  pages = {L012003},
  numpages = {8},
  year = {2022},
  month = {Jan},
  publisher = {American Physical Society},
  doi = {10.1103/PhysRevResearch.4.L012003},
  url = {https://link.aps.org/doi/10.1103/PhysRevResearch.4.L012003}
}

@article{SerbynEtAlQuench2014,
  title = "{Quantum quenches in the many-body localized phase}",
  author = {Serbyn, M. and Papi{\'c}, Z. and Abanin, D.~A.},
  journal = {Phys. Rev. B},
  volume = {90},
  pages = {174302},
  year = {2014},
  doi = {10.1103/PhysRevB.90.174302}
}

@article{Lukin2019_Science,
  title = "{Probing entanglement in a many-body–localized system}",
  author = {Lukin, A. and Rispoli, M. and Schittko, R. and Tai, M. E. and Kaufman, A. M. and Choi, S. and Khemani, V. and L\'{e}onard, J. and Greiner, M.},
  journal = {Science},
  volume = {364},
  number = {6437},
  pages = {256--260},
  year = {2019},
  doi = {10.1126/science.aau0818}
}

@article{AbaninEtAlRMP2019,
  title = "{Colloquium: Many-body localization, thermalization, and entanglement}",
  author = {Abanin, D.~A. and Altman, E. and Bloch, I. and Serbyn, M.},
  journal = {Rev. Mod. Phys.},
  volume = {91},
  pages = {021001},
  year = {2019},
  doi = {10.1103/RevModPhys.91.021001}
}

@Article{10.21468/SciPostPhysCore.4.2.010,
	title={{The Folded Spin-1/2 XXZ Model: I. Diagonalisation, Jamming, and Ground State Properties}},
	author={Lenart Zadnik and Maurizio Fagotti},
	journal={SciPost Phys. Core},
	volume={4},
	pages={010},
	year={2021},
	publisher={SciPost},
	doi={10.21468/SciPostPhysCore.4.2.010},
	url={https://scipost.org/10.21468/SciPostPhysCore.4.2.010},
}

@Article{10.21468/SciPostPhys.10.5.099,
	title={{The folded spin-1/2 XXZ model: II. Thermodynamics and hydrodynamics with a minimal set of charges}},
	author={Lenart Zadnik and Kemal Bidzhiev and Maurizio Fagotti},
	journal={SciPost Phys.},
	volume={10},
	pages={099},
	year={2021},
	publisher={SciPost},
	doi={10.21468/SciPostPhys.10.5.099},
	url={https://scipost.org/10.21468/SciPostPhys.10.5.099},
}

@article{PhysRevE.104.044106,
  title = "{Integrable spin chain with Hilbert space fragmentation and solvable real-time dynamics}",
  author = {Pozsgay, Bal\'azs and Gombor, Tam\'as and Hutsalyuk, Arthur and Jiang, Yunfeng and Pristy\'ak, Levente and Vernier, Eric},
  journal = {Phys. Rev. E},
  volume = {104},
  issue = {4},
  pages = {044106},
  numpages = {27},
  year = {2021},
  month = {Oct},
  publisher = {American Physical Society},
  doi = {10.1103/PhysRevE.104.044106},
  url = {https://link.aps.org/doi/10.1103/PhysRevE.104.044106}
}

@article{SerbynEtAlPRL2013,
  title = "{Universal slow growth of entanglement in interacting strongly disordered systems}",
  author = {Serbyn, M. and Papi{\'c}, Z. and Abanin, D.~A.},
  journal = {Phys. Rev. Lett.},
  volume = {110},
  pages = {260601},
  year = {2013},
  doi = {10.1103/PhysRevLett.110.260601}
}

@article{MorningstarHusePRB2019,
  title = "{Renormalization-group study of the many-body localization transition in one dimension}",
  author = {Morningstar, A. and Huse, D.~A.},
  journal = {Phys. Rev. B},
  volume = {99},
  pages = {224205},
  year = {2019},
  doi = {10.1103/PhysRevB.99.224205}
}

@article{PhysRevB.105.174205,
  title = "{Avalanches and many-body resonances in many-body localized systems}",
  author = {Morningstar, Alan and Colmenarez, Luis and Khemani, Vedika and Luitz, David J. and Huse, David A.},
  journal = {Phys. Rev. B},
  volume = {105},
  issue = {17},
  pages = {174205},
  numpages = {20},
  year = {2022},
  month = {May},
  publisher = {American Physical Society},
  doi = {10.1103/PhysRevB.105.174205},
  url = {https://link.aps.org/doi/10.1103/PhysRevB.105.174205}
}

@article{PhysRevB.93.060201,
  title = "{Extended slow dynamical regime close to the many-body localization transition}",
  author = {Luitz, David J. and Laflorencie, Nicolas and Alet, Fabien},
  journal = {Phys. Rev. B},
  volume = {93},
  issue = {6},
  pages = {060201},
  numpages = {5},
  year = {2016},
  month = {Feb},
  publisher = {American Physical Society},
  doi = {10.1103/PhysRevB.93.060201},
  url = {https://link.aps.org/doi/10.1103/PhysRevB.93.060201}
}

@article{OganesyanHuse2007,
  title = "{Localization of interacting fermions at high temperature}",
  author = {Oganesyan, V. and Huse, D.~A.},
  journal = {Phys. Rev. B},
  volume = {75},
  pages = {155111},
  year = {2007},
  doi = {10.1103/PhysRevB.75.155111}
}

@article{LuitzLaflorencieAletPRB2015,
  title = "{Many-body localization edge in the random-field Heisenberg chain}",
  author = {Luitz, D.~J. and Laflorencie, N. and Alet, F.},
  journal = {Phys. Rev. B},
  volume = {91},
  pages = {081103},
  year = {2015},
  doi = {10.1103/PhysRevB.91.081103}
}

@article{PhysRevLett.118.196801,
  title = {Density Propagator for Many-Body Localization: Finite-Size Effects, Transient Subdiffusion, and Exponential Decay},
  author = {Bera, Soumya and De Tomasi, Giuseppe and Weiner, Felix and Evers, Ferdinand},
  journal = {Phys. Rev. Lett.},
  volume = {118},
  issue = {19},
  pages = {196801},
  numpages = {6},
  year = {2017},
  month = {May},
  publisher = {American Physical Society},
  doi = {10.1103/PhysRevLett.118.196801},
  url = {https://link.aps.org/doi/10.1103/PhysRevLett.118.196801}
}

@article{NandkishoreHuseARCM2015,
  title = "{Many-body localization and thermalization in quantum statistical mechanics}",
  author = {Nandkishore, R. and Huse, D.~A.},
  journal = {Annu. Rev. Condens. Matter Phys.},
  volume = {6},
  pages = {15--38},
  year = {2015},
  doi = {10.1146/annurev-conmatphys-031214-014726}
}

@article{Dabholkar:2024jll,
    author = "Dabholkar, Bhupen and Alet, Fabien",
    title = "{Ergodic and non-ergodic properties of disordered SU(3) chains}",
    eprint = "2403.00442",
    archivePrefix = "arXiv",
    primaryClass = "cond-mat.dis-nn",
    month = "3",
    year = "2024"
}

@article{Bahri2015,
    author = "Bahri, Yasaman and Vosk, Ronen and Altman, Ehud and Vishwanath, Ashvin",
    title = "{Localization and topology protected quantum coherence at the edge of hot matter}",
    journal = {Nature Comm.},
    volume = {91},
    pages = {7341},
    year = {2015},
    doi = {https://doi.org/10.1038/ncomms8341},
    url = {https://arxiv.org/abs/2403.00442}
}

@misc{Miranda2025LargeDI,
      title="{Large deviations in the many-body localization transition: The case of the random-field XXZ chain}", 
      author={Greivin Alfaro Miranda and Fabien Alet and Giulio Biroli and Leticia F. Cugliandolo and Nicolas Laflorencie and Marco Tarzia},
      year={2025},
      eprint={2510.18545},
      archivePrefix={arXiv},
      primaryClass={cond-mat.dis-nn},
      url={https://arxiv.org/abs/2510.18545}, 
}

@article{PhysRevX.4.011052,
  title = "{Hilbert-Glass Transition: New Universality of Temperature-Tuned Many-Body Dynamical Quantum Criticality}",
  author = {Pekker, David and Refael, Gil and Altman, Ehud and Demler, Eugene and Oganesyan, Vadim},
  journal = {Phys. Rev. X},
  volume = {4},
  issue = {1},
  pages = {011052},
  numpages = {12},
  year = {2014},
  month = {Mar},
  publisher = {American Physical Society},
  doi = {10.1103/PhysRevX.4.011052},
  url = {https://link.aps.org/doi/10.1103/PhysRevX.4.011052}
}

@article{ImbrieJSP2016,
  title = "{On many-body localization for quantum spin chains}",
  author = {Imbrie, J.~Z.},
  journal = {J. Stat. Phys.},
  volume = {163},
  pages = {998--1048},
  year = {2016},
  doi = {10.1007/s10955-016-1508-x}
}

@article{ImbrieArXiv2016,
  title = "{Diagonalization and many-body localization for a disordered quantum spin chain}",
  author = {Imbrie, J.~Z.},
  journal = {Phys. Rev. Lett.},
  volume = {117},
  pages = {027201},
  year = {2016},
  doi = {10.1103/PhysRevLett.117.027201}
}

@article{SerbynPapicAbaninPRL2013,
  title = "{Local conservation laws and the structure of the many-body localized states}",
  author = {Serbyn, M. and Papi{\'c}, Z. and Abanin, D.~A.},
  journal = {Phys. Rev. Lett.},
  volume = {111},
  pages = {127201},
  year = {2013},
  doi = {10.1103/PhysRevLett.111.127201}
}

@article{HuseNandkishoreOganesyanPalPRL2013,
  title = "{Localization-protected quantum order}",
  author = {Huse, D.~A. and Nandkishore, R. and Oganesyan, V. and Pal, A. and Sondhi, S.~L.},
  journal = {Phys. Rev. B},
  volume = {88},
  pages = {014206},
  year = {2013},
  doi = {10.1103/PhysRevB.88.014206}
}

@article{ChoiScience2016,
  title = "{Exploring the many-body localization transition in two dimensions}",
  author = {Choi, J.-y. and Hild, S. and Zeiher, J. and Schau{\ss}, P. and Rubio-Abadal, A. and Yefsah, T. and Khemani, V. and Huse, D.~A. and Bloch, I. and Gross, C.},
  journal = {Science},
  volume = {352},
  pages = {1547--1552},
  year = {2016},
  doi = {10.1126/science.aaf8834}
}

@article{SierantEtAlReview2024, 
 title="{Many-body localization in the age of classical computing}", 
 volume={88}, 
 ISSN={1361-6633}, 
 url={10.1088/1361-6633/ad9756}, 
 DOI={10.1088/1361-6633/ad9756}, 
 number={2}, 
 journal={Reports on Progress in Physics}, 
 publisher={IOP Publishing}, 
 author={Sierant, Piotr and Lewenstein, Maciej and Scardicchio, Antonello and Vidmar, Lev and Zakrzewski, Jakub}, 
 year={2025}, 
 month=jan, 
 pages={026502} 
 }

@article{AletLaflorencieReview2018,
  title = "{Many-body localization: An introduction and selected topics}",
  author = {Alet, F. and Laflorencie, N.},
  journal = {C. R. Phys.},
  volume = {19},
  pages = {498--525},
  year = {2018},
  doi = {10.1016/j.crhy.2018.03.003}
}

@article{BardarsonPollmannMoore,
  title = "{Unbounded Growth of Entanglement in Models of Many-Body Localization}",
  author = {Bardarson, Jens H. and Pollmann, Frank and Moore, Joel E.},
  journal = {Phys. Rev. Lett.},
  volume = {109},
  issue = {1},
  pages = {017202},
  numpages = {5},
  year = {2012},
  month = {Jul},
  publisher = {American Physical Society},
  doi = {10.1103/PhysRevLett.109.017202},
  url = {https://link.aps.org/doi/10.1103/PhysRevLett.109.017202}
}

@article{VoskAltmanHusePRL2015,
  title = "{Theory of the many-body localization transition in one-dimensional systems}",
  author = {Vosk, R. and Huse, D.~A. and Altman, E.},
  journal = {Phys. Rev. X},
  volume = {5},
  pages = {031032},
  year = {2015},
  doi = {10.1103/PhysRevX.5.031032}
}

@article{AgarwalAltmanAbanin2017,
  title = "{Rare-region effects and dynamics near the many-body localization transition}",
  author = {Agarwal, K. and Altman, E. and Demler, E. and Gopalakrishnan, S. and Huse, D.~A. and Knap, M.},
  journal = {Ann. Phys.},
  volume = {529},
  pages = {1600326},
  year = {2017},
  doi = {10.1002/andp.201600326}
}

@article{PhysRevLett.95.206603,
  title = "{Interacting Electrons in Disordered Wires: Anderson Localization and Low-$T$ Transport}",
  author = {Gornyi, I. V. and Mirlin, A. D. and Polyakov, D. G.},
  journal = {Phys. Rev. Lett.},
  volume = {95},
  issue = {20},
  pages = {206603},
  numpages = {4},
  year = {2005},
  month = {Nov},
  publisher = {American Physical Society},
  doi = {10.1103/PhysRevLett.95.206603},
  url = {https://link.aps.org/doi/10.1103/PhysRevLett.95.206603}
}

@article{BaskoAleinerAltshuler2006,
  title = "{Metal-insulator transition in a weakly interacting many-electron system with localized single-particle states}",
  author = {Basko, D.~M. and Aleiner, I.~L. and Altshuler, B.~L.},
  journal = {Ann. Phys.},
  volume = {321},
  pages = {1126--1205},
  year = {2006},
  doi = {10.1016/j.aop.2005.11.014}
}

@article{PhysRevB.76.052203,
  title = "{Possible experimental manifestations of the many-body localization}",
  author = {Basko, D. M. and Aleiner, I. L. and Altshuler, B. L.},
  journal = {Phys. Rev. B},
  volume = {76},
  issue = {5},
  pages = {052203},
  numpages = {4},
  year = {2007},
  month = {Aug},
  publisher = {American Physical Society},
  doi = {10.1103/PhysRevB.76.052203},
  url = {https://link.aps.org/doi/10.1103/PhysRevB.76.052203}
}

@article{BaskoAltshuler,
  title = "{On the problem of many-body localization}",
  author = {Basko, D. M. and Aleiner, I. L. and Altshuler, B. L.},
  url = {https://arxiv.org/abs/cond-mat/0602510},
  doi = {10.48550/arXiv.cond-mat/0602510}
}

@article{DumitrescuEtAl2018,
  title = "{Kosterlitz-Thouless scaling at many-body localization phase transitions}",
  author = {Dumitrescu, Philipp T. and Goremykina, Anna and Parameswaran, Siddharth A. and Serbyn, Maksym and Vasseur, Romain},
  journal = {Phys. Rev. B},
  volume = {99},
  issue = {9},
  pages = {094205},
  numpages = {16},
  year = {2019},
  month = {Mar},
  publisher = {American Physical Society},
  doi = {10.1103/PhysRevB.99.094205},
  url = {https://link.aps.org/doi/10.1103/PhysRevB.99.094205}
}

@article{lydzba2024local,
  title={Local integrals of motion in dipole-conserving models with Hilbert space fragmentation},
  author={{\L}yd{\.z}ba, Patrycja and Prelov{\v{s}}ek, Peter and Mierzejewski, Marcin},
  journal={Physical Review Letters},
  volume={132},
  number={22},
  pages={220405},
  year={2024},
  publisher={APS},
  doi = {https://doi.org/10.1103/PhysRevLett.132.220405},
  url = {https://journals.aps.org/prl/abstract/10.1103/PhysRevLett.132.220405}
  
}

@article{classen2025universal,
  title={Universal freezing transitions of dipole-conserving chains},
  author={Classen-Howes, Jonathan and Senese, Riccardo and Prakash, Abhishodh},
  journal={Physical Review B},
  volume={112},
  number={12},
  pages={125148},
  year={2025},
  publisher={APS},
  doi = {https://doi.org/10.1103/2h1v-yx5l},
  url = {https://journals.aps.org/prb/abstract/10.1103/2h1v-yx5l}
}

@article{hart2024exact,
  title="{Exact Mazur bounds in the pair-flip model and beyond}",
  author={Hart, Oliver},
  journal={SciPost Physics Core},
  volume={7},
  number={3},
  pages={040},
  year={2024},
  doi = {10.21468/SciPostPhysCore.7.3.040},
  url = {https://www.scipost.org/10.21468/SciPostPhysCore.7.3.040}
}

@article{stahl2024topologically,
  title="{Topologically stable ergodicity breaking from emergent higher-form symmetries in generalized quantum loop models}",
  author={Stahl, Charles and Nandkishore, Rahul and Hart, Oliver},
  journal={SciPost Physics},
  volume={16},
  number={3},
  pages={068},
  year={2024},
  doi = {10.21468/SciPostPhys.16.3.068},
  url = {https://www.scipost.org/SciPostPhys.16.3.068?acad_field_slug=politicalscience}
}

@article{gp2w-mlkk,
  title = "{Strong Hilbert space fragmentation and fractons from subsystem and higher-form symmetries}",
  author = {Stahl, Charles and Hart, Oliver and Khudorozhkov, Alexey and Nandkishore, Rahul},
  journal = {Phys. Rev. B},
  volume = {112},
  issue = {10},
  pages = {104316},
  numpages = {9},
  year = {2025},
  month = {Sep},
  publisher = {American Physical Society},
  doi = {10.1103/gp2w-mlkk},
  url = {https://link.aps.org/doi/10.1103/gp2w-mlkk}
}

@article{Scardicchio_NN, 
    title="{Foundation neural-networks quantum states as a unified Ansatz for multiple hamiltonians}", 
    volume={16}, 
    ISSN={2041-1723}, 
    url={http://dx.doi.org/10.1038/s41467-025-62098-x}, 
    number={1}, 
    journal={Nature Comm.}, 
    publisher={Springer Science and Business Media LLC}, 
    author={Rende, Riccardo and Viteritti, Luciano Loris and Becca, Federico and Scardicchio, Antonello and Laio, Alessandro and Carleo, Giuseppe}, 
    year={2025}, 
    month=aug 
}

@article{PhysRevB.102.125134,
  title = "{Many-body localization near the critical point}",
  author = {Morningstar, Alan and Huse, David A. and Imbrie, John Z.},
  journal = {Phys. Rev. B},
  volume = {102},
  issue = {12},
  pages = {125134},
  numpages = {10},
  year = {2020},
  month = {Sep},
  publisher = {American Physical Society},
  doi = {10.1103/PhysRevB.102.125134},
  url = {https://link.aps.org/doi/10.1103/PhysRevB.102.125134}
}
\end{document}